\documentclass[english]{article}
\usepackage[T1]{fontenc}
\usepackage[latin9]{inputenc}
\usepackage{geometry}
\usepackage{color}
\usepackage{mathtools}
\usepackage{amsmath}
\usepackage{amssymb}
\usepackage{esint}
\usepackage{babel}
\begin{document}

\title{\textcolor{blue}{\huge{}Semiclassical Newtonian Field Theories Based
On Stochastic Mechanics I}}

\author{Maaneli Derakhshani\thanks{Email: maanelid@yahoo.com and m.derakhshani@uu.nl}}

\maketitle
\begin{center}
\emph{\large{}Institute for History and Foundations of Science \&
Department of Mathematics, Utrecht University, Utrecht, The Netherlands}\emph{}\footnote{Address when this work was initiated: Department of Physics \& Astronomy,
University of Nebraska-Lincoln, Lincoln, NE 68588, USA.}
\par\end{center}

~
\begin{abstract}
This is the first in a two-part series in which we extend non-relativistic
stochastic mechanics, in the ZSM formulation \cite{Derakhshani2016a,Derakhshani2016b},
to semiclassical Newtonian gravity (ZSM-Newton) and semiclassical
Newtonian electrodynamics (ZSM-Coulomb), under the assumption that
the gravitational and electromagnetic fields are fundamentally classical
(i.e., not independently quantized fields). Our key findings are:
(1) a derivation of the usual $N$-particle Schr{\"o}dinger equation
for many particles interacting through operator-valued gravitational
or Coulomb potentials, and (2) recovery of the `single-body' Schr{\"o}dinger-Newton
and Schr{\"o}dinger-Coulomb equations as mean-field equations valid for systems of gravitationally and electrostatically interacting identical particles, respectively, in the weak-coupling large N limit.
We also compare ZSM-Newton/Coulomb to semiclassical Newtonian gravity/electrodynamics
approaches based on standard quantum theory, dynamical collapse theories,
and the de Broglie-Bohm theory. 
\end{abstract}

\section{Introduction}

Semiclassical theories \footnote{As in field theories where the matter sector is described within the
framework of quantum mechanics or quantum field theory, and the gravity
sector is described by classical (c-number) fields.} of gravity and electrodynamics, based on the formalism of standard
quantum theory, have been thoroughly studied over the past 55 years
\cite{Moeller1962,Rosenfeld1963,KibbleRandjbarDaemi1980,PageGeilker1981,HartleHorowitz1981,BirrellDavies1982,Ford1982,Wald1994,Visser2002,Ford2005,HuRouraVerdaguer2004,Hu2008,Kiefer2012,Hu2014,AnastopoulosHu2014,Barut1988a,Barut1988b,BarutDowling1990a,BarutDowling1990b,Milonni1994}.
In the past 20 years or so, semiclassical Newtonian gravity based
on the Schr{\"o}dinger-Newton (SN) equation \cite{Diosi1984,Penrose1996,Penrose1998,Marshall2003,Guzman2003,Salzman2005,Diosi2007,Adler2007,Carlip2008,vanWezel2008,Boughn2009,Meter2011,Giulini2011,Kiefer2012,Salcedo2012,Giulini2012,Bassi2013,Giulini2013,Yang2013,AnastopoulosHu2014,Hu2014,Colin2014,Giulini2014,BahramiBassi2014a,Anastopoulos2015,Grossardt2015,Bahrami2015,Bera2015,DerakProbingGravCat2016,DerAnasHu2016,Grossardt2016,Pino2016,Gan2016,Diosi2016,DerakNewtLimStoGra2017}
has become a popular focus of discussions in the foundations of quantum
mechanics \cite{Penrose1996,Penrose1998,Adler2007,vanWezel2008,Bassi2013,Hu2014,Derakhshani2014,AnastopoulosHu2014,BahramiBassi2014a,Bahrami2015,Diosi2016,DerakNewtLimStoGra2017},
quantum gravity phenomenology \cite{Penrose1996,Penrose1998,Marshall2003,Salzman2005,Carlip2008,Boughn2009,Bassi2013,Giulini2013,Yang2013,BahramiBassi2014a,Giulini2014,Colin2014,Bahrami2015,Anastopoulos2015,Bera2015,Grossardt2015,Grossardt2016,Gan2016,DerakProbingGravCat2016,DerakNewtLimStoGra2017},
and state-of-the-art AMO experimental physics \cite{Marshall2003,RomeroIsart2012,Yang2013,Grossardt2015,Bahrami2015,Grossardt2016,Gan2016,Pino2016,DerAnaHu2016,DerakProbingGravCat2016}.
Variants of the SN equation, based on alternative formulations of
quantum theory, have also been developed \cite{Prezhdo2001,Diosi2007,Landau2012,Diosi2012A,Diosi2012B,Adler2013,Bassi2013,Derakhshani2014,Kafri2014,Nimmrichter2015,Struyve2015,Tilloy2016,DerakProbingGravCat2016},
mostly in the context of dynamical collapse theories \cite{Adler2013,Derakhshani2014,DerakProbingGravCat2016,Diosi2007,Diosi2012A,Diosi2012B,Nimmrichter2015,Landau2012,Tilloy2016}.
Less discussion has been given to the possibility of semiclassical
theories of gravity/electrodynamics based on `hidden-variables' \footnote{This phrase is somewhat misleading for the theories in question, but
we will use it abusingly due to its already widespread use in the
literature.} theories; the only instances we know of are Struyve \cite{Struyve2015},
Kiessling \cite{Kiessling2006}, and Prezhdo-Brooksby \cite{Prezhdo2001}
in the context of the de Broglie-Bohm (dBB) pilot-wave theory \cite{HollandBook1993,Goldstein2013,Duerr2009,DGZbook2012}.
Until now, no such discussion has been given in the context of stochastic
mechanical hidden-variables theories \cite{Nelson1967,Nelson1966,Nelson1985,Bohm1989,Kyprianidis1992,Derakhshani2016a,Derakhshani2016b}. 

A central reason for considering formulations of semiclassical gravity
based on alternative quantum theories is that the SN equation, whether
understood as a mean-field approximation to the standard exact quantum
description of matter-gravity coupling \cite{HartleHorowitz1981,HuRouraVerdaguer2004,Hu2008,Kiefer2012,Hu2014,AnastopoulosHu2014,DerakNewtLimStoGra2017}
or as a fundamental theory describing matter-gravity coupling consistent
with standard quantum theory \cite{KibbleRandjbarDaemi1980,PageGeilker1981,Adler2007,Kiefer2012,Derakhshani2014,AnastopoulosHu2014,BahramiBassi2014a,Giulini2014,Bahrami2015,Bera2015,DerakProbingGravCat2016},
is either very limited in applicability or fatally problematic \cite{PageGeilker1981,HuRouraVerdaguer2004,Ford2005,Adler2007,Hu2008,vanWezel2008,AnastopoulosHu2014,Hu2014,Derakhshani2014,Anastopoulos2015,Diosi2016,DerakNewtLimStoGra2017}. 

Understood as a mean-field theory, the nonlinearity of the SN equation
(or the stochastic SN equation discussed by us in \cite{DerakNewtLimStoGra2017})
means that its solutions lack consistent Born-rule interpretations
\cite{Adler2007,vanWezel2008,Derakhshani2014,Hu2014,Diosi2016} (see
section 4 and subsection 5.1); instead, the SN solutions must be understood
as describing self-gravitating \emph{classical matter fields} that
approximate quantum systems involving large numbers of identical particles
that weakly interact \footnote{In the sense of the coupling scaling as $1/N$, where N is the number
of particles.} quantum-gravitationally \cite{BardosGolseMauser2000,BardosErdosGolseMauserYau2002,Golse2003,Hu2014,AnastopoulosHu2014,DerakNewtLimStoGra2017}.
Moreover, only SN solutions with `small quantum fluctuations' (i.e.,
solutions which don't correspond to superpositions of effectively
orthogonal classical field states) can have this physical interpretation
\cite{Ford1982,Diosi1984,Ford2005,Derakhshani2014,Anastopoulos2015,DerakProbingGravCat2016,DerakNewtLimStoGra2017},
implying that the vast majority of SN solutions are (physically) superfluous. 

Understood as a fundamental theory, the nonlinearity of the SN equation
is fatal because the consequent lack of consistent Born-rule interpretations
for the SN solutions destroys the standard quantum interpretation
of the matter sector of fundamentally-semiclassical gravity based
the SN equation (see subsection 5.1). A fundamentally-semiclassical
description of matter-gravity coupling, based on the SN equation,
would actually be a nonlinear classical field theory that makes empirical
predictions (such as macroscopic semiclassical gravitational cat states;
see section 4 for an example) grossly inconsistent with standard quantum
mechanics and the world of lived experience \cite{Ford1982,Diosi1984,Derakhshani2014,Ford2005,BahramiBassi2014a,DerakProbingGravCat2016}.
(Analogous comments apply to semiclassical electrodynamics based on
the Schr{\"o}dinger-Coulomb (SC) equation \cite{Barut1988a,Barut1988b};
see subsection 5.1 for a discussion.) 

Another key motivation for considering formulations of semiclassical
gravity based on alternative quantum theories is that while the standard
exact quantum description of matter-gravity coupling yields semiclassical
gravity as a consistent mean-field approximation, the matter sector
of the standard exact quantum description is afflicted by the quantum
measurement problem \cite{BellAgainstMeasure,HollandBook1993,Bassi2000,AdlerMP2003,Schlosshauer2004a,Duerr2009,Bassi2013}.
This puts a fundamental limitation on the domain of applicability
of the standard exact quantum description (whether at the Newtonian
level or the fully relativistic level), hence a fundamental limitation
on the domain of applicability of semiclassical gravity (whether at
the Newtonian level or the fully relativistic level). Namely, the
standard exact quantum description and the mean-field semiclassical-gravitational
description are only applicable to laboratory experiments involving
the coupling of gravity to quantum matter, since laboratory experiments
are the only places where the standard quantum formalism can be sensibly
applied. 

Thus it stands to reason that a formulation of quantum theory convincingly
free of the measurement problem might, when extended to a semiclassical
description of gravity (whether as a fundamental theory or a mean-field
theory), yield a superior formulation of semiclassical gravity than
the options based on standard quantum theory. Arguably, this suggestion
has already been confirmed (at least at the Newtonian level) by dynamical
collapse versions of fundamentally-semiclassical Newtonian gravity,
insofar as the models of Derakhshani \cite{Derakhshani2014,DerakProbingGravCat2016}
and Tilloy-Di{\'o}si \cite{Tilloy2016} seem to have consistent statistical
interpretations while adequately suppressing gravitational cat state
solutions at the macroscopic scale. In addition, the works of Struyve
\cite{Struyve2015} and Prezhdo-Brooksby \cite{Prezhdo2001} suggest
that the dBB theory offers a more empirically accurate semiclassical
approximation scheme than does standard quantum theory, at least for
simple examples considered at the relativistic level \cite{Struyve2015}
and the Newtonian level \cite{Prezhdo2001,Struyve2015} (see subsection
5.2 for more detail); however, Struyve has pointed out \cite{Struyve2015}
that the dBB theory does not yield a consistent model of fundamentally-semiclassical
Einstein gravity \footnote{The reason, basically, is that a classical stress-energy tensor built
out of a dBB field beable $\phi_{B}(\mathbf{x},t)$ entails non-conservation
of stress-energy, i.e., $\nabla^{\mu}T_{\mu\nu}\left(\phi_{B}\right)\neq0$,
since $\phi_{B}$ does not conserve total energy along its space-time
trajectory. Thus $T_{\mu\nu}\left(\phi_{B}\right)$ cannot be used
on the right-hand-side of the classical Einstein equations as this
would violate the Bianchi identities $\nabla^{\mu}G_{\mu\nu}=0$.}. In our assessment (see subsection 5.2) the dBB theory does not allow
for a consistent model of fundamentally-semiclassical Newtonian gravity/electrodynamics,
either. It would seem, then, that there does not yet exist a compelling
and widely applicable model of semiclassical gravity based on a theory
of hidden-variables, whether in the form of a fundamental theory of
matter-gravity coupling or a mean-field approximation to an exact
`quantum' description of matter-gravity coupling.

The primary objectives of this two-part series are: (i) to construct
a fundamentally-semiclassical theory of Newtonian gravity/electrodynamics
within the framework of stochastic mechanics, in particular a new
formulation of stochastic mechanics we have recently proposed \cite{Derakhshani2016a,Derakhshani2016b}
to answer the long-standing ``Wallstrom criticism''; (ii) to show
that fundamentally-semiclassical Newtonian gravity/electrodynamics
based on our new formulation of stochastic mechanics - which we call
``zitterbewegung stochastic mechanics'' (ZSM), hence `ZSM-Newton'
and `ZSM-Coulomb' - has a consistent statistical interpretation and
recovers the standard exact quantum description of matter-gravity
coupling as an approximation (valid for all practical purposes), while
also being free of the measurement problem; (iii) to show that the
SN/SC equation and the stochastic SN/SC equation can be recovered
as mean-field approximations for large numbers of identical ZSM particles
that weakly interact \footnote{Also in the sense of coupling scaling as $1/N$.}
classical-gravitationally/electrostatically; and (iv) to show that
ZSM-Newton/Coulomb yields a new `large-N' prescription that makes
it possible to: (a) accurately approximate the time-evolution of a
large number of identical ZSM particles that strongly interact classical-gravitationally/electrostatically,
within a consistent statistical interpretation; (b) avoid macroscopic
semiclassical gravitational cat states and recover classical Newtonian
gravity/electrodynamics for the center-of-mass descriptions of macroscopic
particles; and (c) recover classical Vlasov-Poisson mean-field theory
for macroscopic particles that weakly interact gravitationally/electrostatically. 

In the present paper, we will carry out objectives (i-iii), leaving
(iv) for Part II. We will also compare ZSM-Newton/Coulomb to extant
theories of semiclassical Newtonian gravity/electrodynamics that are
based on standard quantum theory, dynamical collapse theories, and
the de Broglie-Bohm theory.

The paper organization is as follows. Section 2 reviews ZSM for the
case of many free particles. Section 3 formulates the basic equations
of ZSM-Newton/Coulomb, explicates the physical interpretation of those
equations, and shows how the standard exact quantum description of
matter-gravity coupling is recovered as a special case valid for all
practical purposes. Section 4 shows how to recover the SN/SC equation
and the stochastic SN/SC equation as mean-field approximations for
large numbers of identical ZSM particles that weakly interact gravitationally/electrostatically.
Finally, section 5 compares ZSM-Newton/Coulomb to extant theories
of semiclassical Newtonian gravity/electrodynamics, pointing out conceptual
and technical advantages entailed by ZSM-Newton/Coulomb, as well as
possibilities for experimental discrimination between ZSM-Newton/Coulomb
and these other semiclassical theories.

\section{Overview of ZSM for many free particles}

ZSM was developed in order to answer Wallstrom's criticism of stochastic
mechanical reconstructions of Schr{\"o}dinger's equation; namely, that
they don't give a plausible justification for the quantum mechanical
requirement that wavefunctions (for spinless particles) must always
be single-valued while allowing generally multi-valued phases \cite{Wallstrom1989,Wallstrom1994,Derakhshani2016a}.
In other words, why it should be that the wavefunction phase $S$
(in polar form) must change along a closed loop in configuration space
by integer multiples of Planck's constant. A formulation of stochastic
mechanics that plausibly answers this criticism is, in our view, a
necessary condition for seriously considering extensions of stochastic
mechanics to more general physical situations, hence why we will base
our approach on the ZSM formulation.

To prepare for the formulation of ZSM-Newton/Coulomb, it is useful
to first review ZSM for $N$ particles that are classically non-interacting
\cite{Derakhshani2016b}.

Our starting point is the following four phenomenological hypotheses.

First, we posit a Minkowski space-time that contains, on a $t=const$
hypersurface, N harmonic oscillators centered around 3-space positions
$\mathbf{q}_{0i}$ for $i=1,..,N$. As ZSM is a phenomenological framework,
we need not specify here the precise physical nature of these harmonic
oscillators (this is task is left for future work). However, we assume
that these oscillators have, in their respective translational rest
frames, natural frequencies $\omega_{ci}\coloneqq\left(1/\hbar\right)m_{i}c^{2}$,
where $c$ is the speed of light and the $m_{i}$ are mass parameters
that set the scales of the natural frequencies. It is reasonable to
call these natural frequencies ``Compton'' frequencies, hence the
label $\omega_{ci}$. We will refer to these oscillators hereafter
as\emph{ }``zitterbewegung (\emph{zbw}) particles'' \cite{Derakhshani2016a,Derakhshani2016b}. 

Second, we adapt Nelson's ether hypothesis \cite{Nelson1966,Nelson1967nopagelist,Nelson1985nopagelist,Nelson1986,Nelson2005}
by supposing now that the Minkowski space-time is pervaded by a frictionless
classical fluid medium (which we will also call an ``ether''), with
the qualitative properties that (i) it is fluctuating everywhere with
the same intensity, and (ii) it is an oscillating medium with a spectrum
of modes superposed at each point in 3-space. More precisely, we imagine
the ether as a continuous (or effectively continuous) medium composed
of a countably infinite number of fluctuating, stationary, spherical
waves superposed at each point in 3-space, with each wave having a
different fixed angular frequency $\omega_{0}^{i}$ where $i$ denotes
the \emph{i}-th ether mode. The relative phases between the modes
are taken to be random so that each mode is effectively uncorrelated
with every other mode. Again, since ZSM is a phenomenological framework,
specifying the precise physical nature of this ether is left to future
work.

Third, we follow Nelson \cite{Nelson1966,Nelson1967nopagelist,Nelson1985}
in hypothesizing that each particle's center of mass, as a result
of being immersed in the ether, undergoes an approximately \emph{frictionless}
translational Brownian motion (due to the homogeneous and isotropic
ether fluctuations that couple to the particles by possibly electromagnetic,
gravitational, or some other means), as modeled by the first-order
stochastic differential equations

\begin{equation}
d\mathbf{q}_{i}(t)=\mathbf{b}_{i}(q(t),t)dt+d\mathbf{W}_{i}(t).\label{eq:1}
\end{equation}
Here the index $i=1,...,N$, the particle trajectories $q(t)=\{\mathbf{q}_{1}(t),\mathbf{q}_{2}(t),...,\mathbf{q}_{N}(t)\}$
$\in$ $\mathbb{R}^{3N}$, and $\mathbf{W}_{i}(t)$ are Wiener processes
modeling each particle's interaction with the ether fluctuations.
The Wiener increments $d\mathbf{W}_{i}(t)$ are assumed to be Gaussian
with zero mean, independent of $d\mathbf{q}_{i}(s)$ for $s\leq t$,
and with variance

\begin{equation}
E_{t}\left[d\mathbf{W}_{in}(t)d\mathbf{W}_{im}(t)\right]=2\nu_{i}\delta_{nm}dt,\label{eq:2}
\end{equation}
where $E_{t}$ denotes the conditional expectation at time \emph{t}.
We then follow Nelson in hypothesizing that the magnitude of the diffusion
coefficients $\nu_{i}$ are defined by 

\begin{equation}
\nu_{i}\coloneqq\frac{\hbar}{2m_{i}}.\label{eq:3}
\end{equation}
Along with (\ref{eq:1}), we also have the backward equations

\begin{equation}
d\mathbf{q}_{i}(t)=\mathbf{b}_{i*}(q(t),t)+d\mathbf{W}_{i*}(t),\label{eq:4}
\end{equation}
where $\mathbf{b}_{i*}(q(t),t)$ are the mean backward drift velocities,
and $d\mathbf{W}_{i*}(t))$ are the backward Wiener processes. As
in the single-particle case, the $d\mathbf{W}_{i*}(t)$ have all the
properties of $d\mathbf{W}_{i}(t)$ except that they are independent
of the $d\mathbf{q}_{i}(s)$ for $s\geq t$. With these conditions
on $d\mathbf{W}_{i}(t)$ and $d\mathbf{W}_{i*}(t)$, equations (\ref{eq:1})
and (\ref{eq:4}) respectively define forward and backward Markov
processes for $N$ particles on $\mathbb{R}^{3}$ (or, equivalently,
for a single particle on $\mathbb{R}^{3N}$). Note that we take the
$\mathbf{b}_{i}$ $(\mathbf{b}_{i*})$ to be functions of all the
particle positions, $q(t)=\{\mathbf{q}_{1}(t),\mathbf{q}_{2}(t),...,\mathbf{q}_{N}(t)\}$
$\in$ $\mathbb{R}^{3N}$. The reasons are: (i) all the particles
are continuously exchanging energy-momentum with a common background
medium (the ether) and thus are in general physically connected in
their translational motions via $\mathbf{b}_{i}$ $(\mathbf{b}_{i*})$,
insofar as the latter are constrained by the physical properties of
the ether; and (ii) the dynamical equations and initial conditions
for the $\mathbf{b}_{i}$ $(\mathbf{b}_{i*})$ are what will determine
the specific situations under which the latter will be effectively
factorizable functions of the particle positions and when they cannot
be effectively factorized. Hence, at this level, it is only sensible
to write $\mathbf{b}_{i}$ $(\mathbf{b}_{i*})$ as functions of all
the particle positions at a single time. 

Fourth, we suppose that, in their respective (now) instantaneous mean
translational rest frames (IMTRFs), i.e., the frames where $\mathbf{b}_{i}=\mathbf{b}_{i*}=0$,
the \emph{zbw} particles undergo driven oscillations about $\mathbf{q}_{0i}$
by coupling to a narrow band of ether modes that resonantly peak around
their natural frequencies. However, in order that the oscillation
of each \emph{zbw} particle doesn't become unbounded in kinetic energy,
there must be some mechanism by which the \emph{zbw} particles dissipate
energy back into the ether so that, on the average, a steady-state
equilibrium regime is reached for their oscillations. So we posit
that on short relaxation time-scales $\tau_{i}$, which are identical
for \emph{zbw} particles of identical rest masses, the mean energy
absorbed by the \emph{i}-th \emph{zbw} particle, as a result of its
driven oscillation by the resonant ether modes, equals the mean energy
dissipated back to the ether by the \emph{i}-th \emph{zbw} particle.
Accordingly, we suppose that, in this steady-state regime, each particle
undergoes a constant mean harmonic oscillation about its $\mathbf{q}_{0i}$
in its IMTRF, as characterized by the fluctuation-dissipation relation
$<H_{i}>_{steady-state}=\hbar\omega_{ci}=m_{i}c^{2}$ where $<H_{i}>_{steady-state}$
is the conserved mean energy due to the steady-state oscillation of
the \emph{i}-th \emph{zbw} particle. 

Now, as a consequence of this last hypothesis, it follows that in
the IMTRF of the \emph{i}-th particle, the mean \emph{zbw} phase change
is given by

\begin{equation}
\delta\bar{\theta}_{i}^{rest}\coloneqq\omega_{ci}\delta t_{0}=\frac{m_{i}c^{2}}{\hbar}\delta t_{0},\label{eq:5}
\end{equation}
and the corresponding absolute mean phase is

\begin{equation}
\bar{\theta}_{i}^{rest}=\omega_{ci}t_{0}+\phi_{i}=\frac{m_{i}c^{2}}{\hbar}t_{0}+\phi_{i}.\label{eq:6}
\end{equation}
Then the joint (mean) phase for all the particles will just be

\begin{equation}
\bar{\theta}_{joint}^{rest}\coloneqq\sum_{i=1}^{N}\bar{\theta}_{i}=\sum_{i=1}^{N}\left(\omega_{ci}t_{0}+\phi_{i}\right)=\sum_{i=1}^{N}\left(\frac{m_{i}c^{2}}{\hbar}t_{0}+\phi_{i}\right).\label{eq:7}
\end{equation}
Note that we cannot talk of the \emph{zbw} phase other than in the
IMTRFs of the particles, because we cannot transform to a frame in
which $d\mathbf{q}_{i}(t)/dt=0$, as this expression is undefined
for the Wiener process.

Lorentz transforming back to the mean translational lab frame where
$\mathbf{b}_{i}(\mathbf{q}_{i}(t),t)\neq0$ and $\mathbf{b}_{i*}(\mathbf{q}_{i}(t),t)\neq0$,
and approximating the transformation for non-relativistic velocities
so that the gamma factor $\gamma=1/\sqrt{\left(1-b_{i}^{2}/c^{2}\right)}\approx1+b_{i}^{2}/2c^{2},$
it is readily shown \cite{Derakhshani2016b} that the forward and
backward joint phase changes become

\begin{equation}
\delta\bar{\theta}_{joint+}^{lab}(q(t),t)\coloneqq\sum_{i=1}^{N}\frac{\omega_{ci}}{m_{i}c^{2}}\left[E_{i+}(q(t),t)\delta t-m_{i}\mathbf{b}_{i}(q(t),t)\cdot\delta\mathbf{q}_{i}(t)\right]=\frac{1}{\hbar}\left[\sum_{i=1}^{N}E_{i+}(q(t),t)\delta t-\sum_{i=1}^{N}m_{i}\mathbf{b}_{i}(q(t),t)\cdot\delta\mathbf{q}_{i}(t)\right],\label{8}
\end{equation}
and

\begin{equation}
\delta\bar{\theta}_{joint-}^{lab}(q(t),t)\coloneqq\sum_{i=1}^{N}\frac{\omega_{ci}}{m_{i}c^{2}}\left[E_{i-}(q(t),t)\delta t-m_{i}\mathbf{b}_{i*}(q(t),t)\cdot\delta\mathbf{q}_{i}(t)\right]=\frac{1}{\hbar}\left[\sum_{i=1}^{N}E_{i-}(q(t),t)\delta t-\sum_{i=1}^{N}m_{i}\mathbf{b}_{i*}(q(t),t)\cdot\delta\mathbf{q}_{i}(t)\right],\label{9}
\end{equation}
where the forward and backward total particle energies are

\begin{equation}
E_{i+}(q(t),t)\coloneqq m_{i}c^{2}+\frac{1}{2}m_{i}b_{i}^{2},\label{10}
\end{equation}
and

\begin{equation}
E_{i-}(q(t),t)\coloneqq m_{i}c^{2}+\frac{1}{2}m_{i}b_{i*}^{2},\label{11}
\end{equation}
neglecting the momentum term proportional to $b_{i}^{3}/c^{2}$. 

Since each \emph{zbw} particle is essentially a harmonic oscillator,
each has its own well-defined phase along its space-time trajectory.
Consistency with this last statement entails that when $\mathbf{b}_{i}(q,t)\neq\sum_{i}^{N}\mathbf{b}_{i}(\mathbf{q}_{i},t)$,
the joint phase must be a well-defined function of the space-time
trajectories of\emph{ }all the \emph{zbw} particles (since we posit
that all particles remain harmonic oscillators despite having their
oscillations physically coupled through the common ether medium they
interact with). Thus we can see that, for a closed loop \emph{L} along
which each \emph{zbw} particle can be physically or virtually displaced,
we have the constraint

\begin{equation}
\sum_{i=1}^{N}\oint_{L}\delta_{i}\bar{\theta}_{joint+,-}^{lab}=2\pi n,\label{12}
\end{equation}
where $n\in\mathbb{Z}$. Moreover, condition (\ref{12}) will also
hold for a closed loop with time held constant. 

Associated to (1) and (4) in the lab frame are the forward and backward
Fokker-Planck equations

\begin{equation}
\frac{\partial\rho(q,t)}{\partial t}=-\sum_{i=1}^{N}\nabla_{i}\cdot\left[\mathbf{b}_{i}(q,t)\rho(q,t)\right]+\sum_{i=1}^{N}\frac{\hbar}{2m_{i}}\nabla_{i}^{2}\rho(q,t),
\end{equation}
and

\begin{equation}
\frac{\partial\rho(q,t)}{\partial t}=-\sum_{i=1}^{N}\nabla_{i}\cdot\left[\mathbf{b}_{i*}(q,t)\rho(q,t)\right]-\sum_{i=1}^{N}\frac{\hbar}{2m_{i}}\nabla_{i}^{2}\rho(q,t),
\end{equation}
where $\rho(q,t)$ is the probability density for the particle trajectories
and satisfies the normalization condition 
\begin{equation}
\int_{\mathbb{R}^{3N}}\rho_{0}(q)d^{3N}q=1.
\end{equation}
Imposing Nelson's time-symmetric kinematic constraints

\begin{equation}
\mathbf{v}_{i}(q,t)\coloneqq\frac{1}{2}\left[\mathbf{b}_{i}(q,t)+\mathbf{b}_{i*}(q,t)\right]=\frac{\nabla_{i}S(q,t)}{m_{i}}|_{\mathbf{q}_{j}=\mathbf{q}_{j}(t)},
\end{equation}
and

\begin{equation}
\mathbf{u}_{i}(q,t)\coloneqq\frac{1}{2}\left[\mathbf{b}_{i}(q,t)-\mathbf{b}_{i*}(q,t)\right]=\frac{\hbar}{2m_{i}}\frac{\nabla_{i}\rho(q,t)}{\rho(q,t)}|_{\mathbf{q}_{j}=\mathbf{q}_{j}(t)},
\end{equation}
where $i,j=1,...,N$, then (13-14) reduce to

\begin{equation}
\frac{\partial\rho({\normalcolor q},t)}{\partial t}=-\sum_{i=1}^{N}\nabla_{i}\cdot\left[\frac{\nabla_{i}S(q,t)}{m_{i}}\rho(q,t)\right],
\end{equation}
with $\mathbf{b}_{i}=\mathbf{v}_{i}+\mathbf{u}_{i}$ and $\mathbf{b}_{i*}=\mathbf{v}_{i}-\mathbf{u}_{i}$. 

To give (17) a coherent physical interpretation, we introduce the
presence of an external (to the particle) osmotic potential $U(q,t)$
which couples to the $i$-th particle as $R(q(t),t)\coloneqq\mu U(q(t),t)$
(assuming that the coupling constant $\mu$ is identical for particles
of the same species), and imparts a momentum, $\nabla_{i}R(q,t)|_{\mathbf{q}_{j}=\mathbf{q}_{j}(t)}$.
This momentum then gets counter-balanced by the ether fluid's osmotic
impulse pressure, $\left(\hbar/2m_{i}\right)\nabla_{i}ln[n(q,t)]|_{\mathbf{q}_{j}=\mathbf{q}_{j}(t)}$,
so that the $N$-particle osmotic velocity is the equilibrium velocity
acquired by the $i$-th particle when $\nabla_{i}R/m_{i}=\left(\hbar/2m_{i}\right)\nabla_{i}\rho/\rho$
(using $\rho=n/N$), which implies $\rho=e^{2R/\hbar}$ for all times.
It is supposed that $R$ generally depends on the coordinates of all
the other particles because: (i) if $U$ was an independently existing
field on configuration space, rather than sourced by the ether, then
the diffusions of the particles through the ether would not be conservative
(i.e., energy conserving), in contradiction with Nelson's hypothesis
that the diffusions \emph{are} conservative, and (ii) since the particles
continuously exchange energy-momentum with the ether, the functional
dependence of $U$ should be determined by the dynamical coupling
of the ether to the particles as well as the magnitude of the inter-particle
physical interactions (whether through a classical inter-particle
potential or, in the free particle case, just through the ether). 

To obtain the time-symmetric mean dynamics for the translational motions
of the \emph{N} particles, i.e., the dynamics for $S$, we integrate
the time-asymmetric joint phases, (8) and (9), and then average the
two to get 

\begin{equation}
\begin{aligned}\bar{\theta}_{joint}^{lab}(q(t),t) & \coloneqq\sum_{i=1}^{N}\frac{\omega_{ci}}{m_{i}c^{2}}\left[\int E_{i}(q(t),t)dt-\int\frac{m_{i}}{2}\left(\mathbf{b}_{i}(q(t),t)+\mathbf{b}_{i*}(q(t),t)\right)\cdot d\mathbf{q}_{i}(t)\right]+\sum_{i=1}^{N}\phi_{i}\\
 & =\frac{1}{\hbar}\left[\int\sum_{i=1}^{N}E_{i}(q(t),t)dt-\int\sum_{i=1}^{N}\frac{m_{i}}{2}\left(\mathbf{b}_{i}(q(t),t)+\mathbf{b}_{i*}(q(t),t)\right)\cdot d\mathbf{q}_{i}(t)\right]+\sum_{i=1}^{N}\phi_{i},
\end{aligned}
\end{equation}
where, from the kinematic constraints (16-17) we have

\begin{equation}
E_{i}(q(t),t)\coloneqq m_{i}c^{2}+\frac{1}{2}\left[\frac{1}{2}m_{i}b_{i}^{2}+\frac{1}{2}m_{i}b_{i*}^{2}\right]=m_{i}c^{2}+\frac{1}{2}m_{i}v_{i}^{2}+\frac{1}{2}m_{i}u_{i}^{2}.
\end{equation}
It also follows that 

\begin{equation}
\mathbf{p}_{i}(q(t),t)=-\hbar\nabla_{i}\bar{\theta}_{joint}^{lab}(q,t)|_{\mathbf{q}_{j}=\mathbf{q}_{j}(t)}=\nabla_{i}S(q,t)|_{\mathbf{q}_{j}=\mathbf{q}_{j}(t)},
\end{equation}
which establishes the \emph{i}-th Nelsonian current velocity as the
\emph{i}-th translational mean velocity component of (19), and the
velocity potential $S$ as the joint mean phase of the \emph{zbw}
particles undergoing Nelsonian diffusions. 

To construct the dynamics for the \emph{zbw} particles, we first introduce
the mean forward and mean backward derivatives:

\begin{equation}
D\mathbf{q}_{i}(t)\coloneqq\underset{_{\Delta t\rightarrow0^{+}}}{lim}E_{t}\left[\frac{q_{i}(t+\Delta t)-q_{i}(t)}{\Delta t}\right]=\mathbf{b}_{i}(q(t),t),
\end{equation}
and

\begin{equation}
D_{*}\mathbf{q}_{i}(t)\coloneqq\underset{_{\Delta t\rightarrow0^{+}}}{lim}E_{t}\left[\frac{q_{i}(t)-q_{i}(t-\Delta t)}{\Delta t}\right]=\mathbf{b}_{i*}(q(t),t),
\end{equation}
where we used the Gaussianity of $d\mathbf{W}_{i}(t)$ and $d\mathbf{W}_{i*}(t)$
in equations (1) and (4). Finding $D\mathbf{b}_{i}(q(t),t)$ (or $D_{*}\mathbf{b}_{i}(q(t),t)$)
is straightforward: expand $\mathbf{b}_{i}$ in a Taylor series up
to terms of order two in $d\mathbf{q}_{i}(t)$, replace $dq_{i}(t)$
by $dW_{i}(t)$ in the last term, and replace $d\mathbf{q}_{i}(t)\cdot\nabla_{i}\mathbf{b}_{i}|_{\mathbf{q}_{j}=\mathbf{q}_{j}(t)}$
by $\mathbf{b}_{i}(\mathbf{q}(t),t)\cdot\nabla_{i}\mathbf{b}_{i}|_{\mathbf{q}_{j}=\mathbf{q}_{j}(t)}$
when taking the conditional expectation at time \emph{t} (since $d\mathbf{W}_{i}(t)$
is independent of $\mathbf{q}_{i}(t)$ and has mean 0). We then have

\begin{equation}
D\mathbf{b}_{i}(q(t),t)=\left[\frac{\partial}{\partial t}+\sum_{i=1}^{N}\mathbf{b}_{i}(q(t),t)\cdot\nabla_{i}+\sum_{i=1}^{N}\frac{\hbar}{2m_{i}}\nabla_{i}^{2}\right]\mathbf{b}_{i}(q(t),t),
\end{equation}
and

\begin{equation}
D_{*}\mathbf{b}_{i*}(q(t),t)=\left[\frac{\partial}{\partial t}+\sum_{i=1}^{N}\mathbf{b}_{i*}(q(t),t)\cdot\nabla_{i}-\sum_{i=1}^{N}\frac{\hbar}{2m_{i}}\nabla_{i}^{2}\right]\mathbf{b}_{i*}(q(t),t).
\end{equation}

With these derivatives operators in hand, we can define the ensemble-averaged,
time-symmetric particle phase as

\begin{equation}
\begin{aligned}J(q,t)=\int_{\mathbb{R}^{3N}}d^{3N}\mathbf{q}\,\rho(q,t)\,\bar{\theta}_{joint}^{lab}(q(t),t) & =\int_{\mathbb{R}^{3N}}d^{3N}\mathbf{q}\,\rho\sum_{i=1}^{N}\frac{\omega_{ci}}{m_{i}c^{2}}\left[\int_{t_{I}}^{t_{F}}E_{i}dt-\int_{q_{iI}}^{q_{iF}}m_{i}\mathbf{v}_{i}\cdot d\mathbf{q}_{i}(t)\right]+\int_{\mathbb{R}^{3N}}d^{3N}\mathbf{q}\,\rho\sum_{i=1}^{N}\phi_{i}\\
 & =\frac{1}{\hbar}\left(\int_{t_{I}}^{t_{F}}<\sum_{i=1}^{N}E_{i}>dt-\int_{q_{I}}^{q_{F}}<\sum_{i=1}^{N}\mathbf{p}_{i}\cdot d\mathbf{q}_{i}(t)>\right)+\sum_{i=1}^{N}\phi_{i},\\
 & =\int_{\mathbb{R}^{3N}}d^{3N}\mathbf{q}\,\rho\int_{t_{I}}^{t_{F}}\sum_{i=1}^{N}\left\{ m_{i}c^{2}+\frac{1}{2}m_{i}v_{i}^{2}+\frac{1}{2}m_{i}u_{i}^{2}\right\} dt+\sum_{i=1}^{N}\phi_{i},\\
 & =\int_{\mathbb{R}^{3N}}d^{3N}\mathbf{q}\,\rho\int_{t_{I}}^{t_{F}}\sum_{i=1}^{N}\left\{ m_{i}c^{2}+\frac{1}{2}\left[\frac{1}{2}m_{i}\left(D\mathbf{q}_{i}(t)\right)^{2}+\frac{1}{2}m_{i}\left(D_{*}\mathbf{q}_{i}(t)\right)^{2}\right]\right\} dt+\sum_{i=1}^{N}\phi_{i},
\end{aligned}
\end{equation}
where we have used the time-symmetric mean Legendre transformation
\begin{equation}
L_{i}\coloneqq\frac{1}{2}\left[(m\mathbf{b}_{i})\cdot\mathbf{b}_{i}+(m\mathbf{b}_{i*})\cdot\mathbf{b}_{i*}\right]-\frac{1}{2}\left(E_{i+}+E_{i-}\right)=(m\mathbf{v}_{i})\cdot\mathbf{v}_{i}+(m\mathbf{u}_{i})\cdot\mathbf{u}_{i}-E_{i},
\end{equation}
and the relation $\bar{\theta}_{joint}^{lab}=-\frac{1}{\hbar}S$.
Following Yasue \cite{Yasue1981a,Yasue1981b}, we can apply the stochastic
variational principle

\begin{equation}
J(q,t)=extremal,
\end{equation}
which by straightforward computation \cite{Derakhshani2016b} yields 

\begin{equation}
\sum_{i=1}^{N}\frac{m_{i}}{2}\left[D_{*}D+DD_{*}\right]\mathbf{q}_{i}(t)=0.
\end{equation}

Moreover, since by D'Alembert's principle the $\delta\mathbf{q}_{i}(t)$
are independent (see Appendix A of \cite{Derakhshani2016b}), it follows
from (29) that we have the individual ``mean acceleration'' equations
of motion

\begin{equation}
m_{i}\mathbf{a}_{i}(q(t),t)\coloneqq\frac{m_{i}}{2}\left[D_{*}D+DD_{*}\right]\mathbf{q}_{i}(t)=0.
\end{equation}
By applying the mean derivatives in (29), using that $\mathbf{b}_{i}=\mathbf{v}_{i}+\mathbf{u}_{i}$
and $\mathbf{b}_{i*}=\mathbf{v}_{i}-\mathbf{u}_{i}$, and replacing
$q(t)$ with $q$ on both sides, straightforward manipulations give

\begin{equation}
\sum_{i=1}^{N}m_{i}\left[\partial_{t}\mathbf{v}_{i}+\mathbf{v}_{i}\cdot\nabla_{i}\mathbf{v}_{i}-\mathbf{u}_{i}\cdot\nabla_{i}\mathbf{u}_{i}-\frac{\hbar}{2m_{i}}\nabla_{i}^{2}\mathbf{u}_{i}\right]=0,
\end{equation}
Computing the derivatives in (31), we obtain

\begin{equation}
\begin{aligned}\sum_{i=1}^{N}m_{i}\mathbf{a}_{i}(q(t),t) & =\sum_{i=1}^{N}m_{i}\left[\frac{\partial\mathbf{v}_{i}(q,t)}{\partial t}+\mathbf{v}_{i}(q,t)\cdot\nabla_{i}\mathbf{v}_{i}(q,t)-\mathbf{u}_{i}(q,t)\cdot\nabla_{i}\mathbf{u}_{i}(q,t)-\frac{\hbar}{2m_{i}}\nabla_{i}^{2}\mathbf{u}_{i}(q,t)\right]|_{\mathbf{q}_{j}=\mathbf{q}_{j}(t)}\\
 & =\sum_{i=1}^{N}\nabla_{i}\left[\frac{\partial S(q,t)}{\partial t}+\frac{\left(\nabla_{i}S(q,t)\right)^{2}}{2m_{i}}-\frac{\hbar^{2}}{2m_{i}}\frac{\nabla_{i}^{2}\sqrt{\rho(q,t)}}{\sqrt{\rho(q,t)}}\right]|_{\mathbf{q}_{j}=\mathbf{q}_{j}(t)}=0,
\end{aligned}
\end{equation}
Integrating both sides of (32), setting the arbitrary integration
constants equal to the rest energies, and replacing $q(t)$ with $q$,
we then have the \emph{N}-particle quantum Hamilton-Jacobi equation

\begin{equation}
-\partial_{t}S(q,t)=\sum_{i=1}^{N}m_{i}c^{2}+\sum_{i=1}^{N}\frac{\left(\nabla_{i}S(q,t)\right)^{2}}{2m_{i}}-\sum_{i=1}^{N}\frac{\hbar^{2}}{2m_{i}}\frac{\nabla_{i}^{2}\sqrt{\rho(q,t)}}{\sqrt{\rho(q,t)}}.
\end{equation}
This equation describes the total energy field over N Gibbsian statistical
ensembles. That is, each coordinate $\mathbf{q}_{i}$ of $S$ is associated
to a fictitious ensemble of identical noninteracting \emph{zbw} particles,
where the members of the ensemble differ from each other only by their
initial positions and velocities; the ensemble density is given by
$\rho$, and reflects ignorance of the actual position and velocity
of the actual $i$-th \emph{zbw} particle at time $t$. 

Upon evaluation at $q=q(t),$ we have the total energy of the actual
particles along their actual mean trajectories. We can now see explicitly
that the local evolution equation for the time-symmetric phase (19),
under the variational constraint (28), will just be (33). 

The general solution of (33), i.e., the joint phase field of the \emph{zbw}
particles in the time-symmetric mean lab frame (the frame in which
the current velocities of the \emph{zbw} particles are non-zero),
is clearly of the form

\begin{equation}
S(q,t)=\sum_{i=1}^{N}\int\mathbf{p}_{i}(q,t)\cdot d\mathbf{q}_{i}-\sum_{i=1}^{N}\int E_{i}(q,t)dt-\sum_{i=1}^{N}\hbar\phi_{i}.
\end{equation}
Since each \emph{zbw} particle is posited to be a harmonic oscillator
of identical type, each has its own well-defined phase along its space-time
trajectory. Consistency with this last statement implies that when
$\mathbf{v}_{i}(q,t)\neq\sum_{i}^{N}\mathbf{v}_{i}(\mathbf{q}_{i},t)$,
the joint mean phase must be a well-defined function of the time-symmetric
mean trajectories of all \emph{zbw} particles (since we posit that
all the \emph{zbw} particles remain harmonic oscillators despite having
their oscillations physically coupled through the common ether medium
they interact with). Then, for a closed loop \emph{L} along which
each \emph{zbw} particle can be physically or virtually displaced,
it follows that 

\begin{equation}
\sum_{i=1}^{N}\oint_{L}\delta_{i}S(q(t),t)=\sum_{i=1}^{N}\oint_{L}\left[\mathbf{p}_{i}(q(t),t)\cdot\delta\mathbf{q}_{i}(t)-E_{i}(q(t),t)\delta t\right]=nh.
\end{equation}
And for a closed loop $L$ with $\delta t=0$, we have

\begin{equation}
\sum_{i=1}^{N}\oint_{L}\delta_{i}S(q(t),t)=\sum_{i=1}^{N}\oint_{L}\mathbf{p}_{i}\cdot\delta\mathbf{q}_{i}(t)=\sum_{i=1}^{N}\oint_{L}\mathbf{\nabla}_{i}S(q,t)|_{\mathbf{q}_{j}=\mathbf{q}_{j}(t)}\cdot\delta\mathbf{q}_{i}(t)=nh.
\end{equation}
If we also consider the joint phase field $S(q,t)$, a field over
the \emph{N} ensembles, then the same physical reasoning applied to
each member of the \emph{i}-th ensemble yields

\begin{equation}
\sum_{i=1}^{N}\oint_{L}d_{i}S(q,t)=\sum_{i=1}^{N}\oint_{L}\mathbf{p}_{i}\cdot d\mathbf{q}_{i}=\sum_{i=1}^{N}\oint_{L}\nabla_{i}S(q,t)\cdot d\mathbf{q}_{i}=nh.
\end{equation}
In other words, the loop integral is now an integral of the ensemble's
momentum field along any closed \textit{mathematical} loop in 3-space
with time held constant; that is, a closed loop around which the actual
(\emph{i}-th) particle with momentum $\mathbf{p}_{i}$ \emph{could
potentially be displaced}, starting from any possible position $\mathbf{q}_{i}$
it can occupy at fixed time $t$. It also clear that (37) implies
phase quantization for each individual particle ensemble, upon keeping
all but the \emph{i}-th coordinate fixed and performing the closed-loop
integration. 

Applying the Madelung transformation \cite{Bohm1952I,Takabayasi1952,Nelson1967nopagelist,Nelson1985nopagelist,Wallstrom1989,HollandBook1993,BohmHiley1993,Wallstrom1994}
to the combination of (18), (33), and (37), we can construct the \emph{N}-particle
Schr{\"o}dinger equation

\begin{equation}
i\hbar\frac{\partial\psi(q,t)}{\partial t}=\sum_{i=1}^{N}\left[-\frac{\hbar^{2}}{2m_{i}}\nabla_{i}^{2}+m_{i}c^{2}\right]\psi(q,t),
\end{equation}
where the \emph{N}-particle wavefunction $\psi(q,t)=\sqrt{\rho(q,t)}e^{iS(q,t)/\hbar}$
is single-valued by (37). Of course, in the case of non-interacting
particles, the $N$-particle wavefunction is nodeless and thus we
will have $n=0$ in (37); only in cases of particles in bound states
can we get $n>0$. 

We should note that the solutions of (38) are generally non-factorizable
fields on 3N-dimensional configuration space, which implies non-separability
of $S$ and $R$ (hence non-factorizability of $\rho$) in general.
Insofar as ZSM starts with the heuristic hypothesis of an ontic ether
that lives in 3-space and couples to ontic \emph{zbw} particles in
3-space, this would seem \emph{prima facie} paradoxical, assuming
one takes the mathematical representation of $S$ and $R$ as a literal
indication of the ontic nature of the hypothesized ether (i.e., that
if $S$ and $R$ live in configuration space, then so must the ether).
As discussed at length in \cite{Derakhshani2016b}, there are three
possible ways to resolve this apparent inconsistency: (i) postulate
that the ether lives in configuration space, but, as a matter of physical
law, determines the motion of \emph{N} \emph{zbw} particles in 3-space;
(ii) postulate that the ether lives in configuration space along with
a \emph{zbw} `world particle' (in analogy with Albert's formulation
of the de Broglie-Bohm theory \cite{Albert2015}), and employ a philosophical
functionalist analysis to deduce the emergence of \emph{N} \emph{zbw}
particles floating in a common 3-space; and (iii) view the $S$ and
$R$ fields on configuration space as convenient mathematical representations
of some corresponding ontic fields on 3-space (in analogy with Norsen's
``TELB'' approach to the de Broglie-Bohm theory \cite{Norsen2010,Norsen2014})
which couple to \emph{N} \emph{zbw} particles in 3-space. As also
discussed in \cite{Derakhshani2016b}, we view option (iii) to be
the most natural and fruitful one for ZSM, and we will implicitly
assume this viewpoint throughout this paper.

With the overview completed, we can now develop ZSM-Newton/Coulomb.

\section{ZSM-Newton/Coulomb: Basic equations}

ZSM-Newton/Coulomb is just the generalization of $N$-particle ZSM
to include classical Newtonian gravitational and Coulomb interactions
between the $zbw$ particles. 

We suppose again that each particle undergoes a mean \emph{zbw} oscillation
in its IMTRF, and now also that each \emph{zbw} particle carries charge
$e_{i}$, making them classical charged harmonic oscillators of identical
type. (We subject these particles to the hypothetical constraint of
no electromagnetic radiation emitted when there is no translational
motion; or the constraint that the oscillation of the charge is radially
symmetric so that there is no net energy radiated; or, if the ether
turns out to be electromagnetic in nature as Nelson suggested \cite{Nelson1985nopagelist,Nelson1986},
then that the steady-state \emph{zbw} oscillations of the particles
are due to a balancing between the time-averaged electromagnetic energy
absorbed via the driven oscillations of the particle charges, and
the time-averaged electromagnetic energy radiated back to the ether
by the particles.) So the classical Newtonian gravitational and Coulomb
interactions between the particles are defined by the gravitational
potential (in CGS units) 
\begin{equation}
V_{g}^{int}(\mathbf{q}_{i}(t),\mathbf{q}_{j}(t))=\sum_{i=1}^{N}\frac{m_{i}\Phi_{g}}{2}=-\sum_{i=1}^{N}\frac{m_{i}}{2}\sum_{j=1}^{N(j\neq i)}\frac{m_{j}}{|\mathbf{q}_{i}(t)-\mathbf{q}_{j}(t)|},
\end{equation}
and the Coulomb potential 
\begin{equation}
V_{c}^{int}(\mathbf{q}_{i}(t),\mathbf{q}_{j}(t))=\sum_{i=1}^{N}\frac{e_{i}\Phi_{c}}{2}=\sum_{i=1}^{N}\frac{e_{i}}{2}\sum_{j=1}^{N(j\neq i)}\frac{e_{j}}{|\mathbf{q}_{i}(t)-\mathbf{q}_{j}(t)|}
\end{equation}
respectively, under the point-like interaction assumption $|\mathbf{q}_{i}(t)-\mathbf{q}_{j}(t)|\gg\lambda_{c}$. 

Then the joint-phase change of the \emph{zbw} particles in the mean
forward joint lab frame ($b_{i}\ll c$ approximated) is defined by

\begin{equation}
\begin{aligned}\delta\bar{\theta}_{joint+}^{lab}(\mathbf{q}_{i}(t),\mathbf{q}_{j}(t),t) & =\left[\sum_{i=1}^{N}\omega_{ic}+\sum_{i=1}^{N}\omega_{ci}\frac{b_{i}^{2}}{2c^{2}}+\sum_{i=1}^{N}\omega_{ci}\left(\frac{\Phi_{g}}{2c^{2}}+\frac{e_{i}\Phi_{c}}{m_{i}c^{2}}\right)\right]\left(\delta t-\sum_{i=1}^{N}\frac{\mathbf{b}_{0i}}{c^{2}}\cdot\delta\mathbf{q}_{i}(t)\right)\\
 & =\sum_{i=1}^{N}\left[\omega_{ic}+\omega_{ci}\frac{b_{i}^{2}}{2c^{2}}+\omega_{ci}\left(\frac{m_{i}\Phi_{g}}{2m_{i}c^{2}}+\frac{e_{i}\Phi_{c}}{m_{i}c^{2}}\right)\right]\delta t-\sum_{i=1}^{N}\omega_{ci}\left(\frac{\mathbf{b}_{i}}{c^{2}}\right)|_{\mathbf{q}_{j}=\mathbf{q}_{j}(t)}\cdot\delta\mathbf{q}_{i}(t)\\
 & =\frac{1}{\hbar}\left[\left(\sum_{i=1}^{N}m_{i}c^{2}+\sum_{i=1}^{N}\frac{m_{i}b_{i}^{2}}{2}+V_{g}^{int}+V_{c}^{int}\right)\delta t-\sum_{i=1}^{N}m_{i}\mathbf{b}_{i}\cdot\delta\mathbf{q}_{i}(t)\right],
\end{aligned}
\end{equation}
where $\delta\theta_{joint-}^{lab}$ differs by the replacement of
$\mathbf{b}_{i}$ with $\mathbf{b}_{i*}$. Note that when $|\mathbf{q}_{i}(t)-\mathbf{q}_{j}(t)|$
becomes sufficiently great that $V_{g,c}^{int}$ is negligible, (41)
reduces to an effectively factorizable sum of the mean forward phase
changes for all the \emph{zbw} particles. (Effectively, because the
ether will of course still physically correlate the phase changes
of the particles, even if negligibly.) We can then write 

\begin{equation}
\delta\bar{\theta}_{joint+}^{lab}(\mathbf{q}_{i}(t),\mathbf{q}_{j}(t),t)=\frac{1}{\hbar}\left[E_{joint+}\delta t-\sum_{i=1}^{N}m_{i}\mathbf{b}_{i}\cdot\delta\mathbf{q}_{i}(t)\right],
\end{equation}
and

\begin{equation}
\delta\bar{\theta}_{joint-}^{lab}(\mathbf{q}_{i}(t),\mathbf{q}_{j}(t),t)=\frac{1}{\hbar}\left[E_{joint-}\delta t-\sum_{i=1}^{N}m_{i}\mathbf{b}_{i*}\cdot\delta\mathbf{q}_{i}(t)\right].
\end{equation}

Because each \emph{zbw} particle is a harmonic oscillator, each has
a well-defined phase along its mean forward/backward space-time trajectory.
Consistency with this last statement entails that when $V_{g,c}^{int}>0$,
the joint phase must be a well-defined function of the mean forward/backward
space-time trajectories of all the \emph{zbw} particles (since we
again posit that all the \emph{zbw }particles remain harmonic oscillators
when coupled to each other via $V_{g,c}^{int}$). Then for a closed
loop \emph{L,} along which each \emph{zbw} particle can be physically
or virtually displaced, the mean forward joint phase in the lab frame
will satisfy

\begin{equation}
\sum_{i=1}^{N}\oint_{L}\delta_{i}\bar{\theta}_{joint+}^{lab}=2\pi n,
\end{equation}
and for a loop with time held fixed

\begin{equation}
\sum_{i=1}^{N}\oint_{L}m_{i}\mathbf{b}_{i}\cdot\delta\mathbf{q}_{i}(t)=nh,
\end{equation}
and likewise for the mean backward joint phase. It also follows from
(44-45) that

\begin{equation}
\oint_{L}\delta_{1}\bar{\theta}_{joint+}^{lab}=2\pi n,
\end{equation}
and

\begin{equation}
\oint_{L}m_{1}\mathbf{b}_{1}\cdot\delta\mathbf{q}_{1}(t)=nh,
\end{equation}
where the closed-loop integral here keeps the coordinates of all the
other particles fixed while particle 1 is displaced along \emph{L}. 

In the lab frame, the forward and backward stochastic differential
equations for the translational motion are then 

\begin{equation}
d\mathbf{q}_{i}(t)=\mathbf{b}_{i}(q(t),t)+d\mathbf{W}_{i}(t),
\end{equation}
and

\begin{equation}
d\mathbf{q}_{i}(t)=\mathbf{b}_{i*}(q(t),t)+d\mathbf{W}_{i*}(t),
\end{equation}
with corresponding Fokker-Planck equations

\begin{equation}
\frac{\partial\rho(q,t)}{\partial t}=-\sum_{i=1}^{N}\nabla_{i}\cdot\left[\left(\mathbf{b}_{i}(q,t)\right)\rho(q,t)\right]+\sum_{i=1}^{N}\frac{\hbar}{2m_{i}}\nabla_{i}^{2}\rho(q,t),
\end{equation}
and

\begin{equation}
\frac{\partial\rho(q,t)}{\partial t}=-\sum_{i=1}^{N}\nabla_{i}\cdot\left[\left(\mathbf{b}_{i*}(q,t)\right)\rho(q,t)\right]-\sum_{i=1}^{N}\frac{\hbar}{2m_{i}}\nabla_{i}^{2}\rho(q,t).
\end{equation}
Imposing the time-symmetric kinematic constraints (16-17), (50-51)
reduce to

\begin{equation}
\frac{\partial\rho}{\partial t}=-\sum_{i=1}^{N}\nabla_{i}\cdot\left[\frac{\nabla_{i}S}{m_{i}}\rho\right],
\end{equation}
and we have again $\mathbf{b}_{i}=\mathbf{v}_{i}+\mathbf{u}_{i}$
and $\mathbf{b}{}_{i*}=\mathbf{v}_{i}-\mathbf{u}_{i}$. 

Now, integrating $\delta\theta_{joint+}^{lab}$ and $\delta\theta_{joint-}^{lab}$,
and then averaging the two, we have

\begin{equation}
\bar{\theta}_{joint}^{lab}\coloneqq\frac{1}{\hbar}\left[\int E_{joint}dt-\sum_{i=1}^{N}\int\frac{m_{i}}{2}\left(\mathbf{b}_{i}+\mathbf{b}{}_{i*}\right)\cdot d\mathbf{q}_{i}(t)\right]+\sum_{i=1}^{N}\phi_{i},
\end{equation}
where

\begin{equation}
\begin{aligned}E_{joint} & \coloneqq\frac{1}{2}\left[E_{joint+}+E_{joint-}\right]=\sum_{i=1}^{N}m_{i}c^{2}+\sum_{i=1}^{N}\frac{1}{2}\left[\frac{1}{2}m_{i}b_{i}{}^{2}+\frac{1}{2}mb_{i*}{}^{2}\right]+V_{g}^{int}+V_{c}^{int}\\
 & =\sum_{i=1}^{N}m_{i}c^{2}+\sum_{i=1}^{N}\left[\frac{1}{2}m_{i}v{}_{i}{}^{2}+\frac{1}{2}m_{i}u_{i}^{2}\right]+V_{g}^{int}+V_{c}^{int}.
\end{aligned}
\end{equation}
Then 

\begin{equation}
\begin{aligned}J(q,t)\coloneqq\int_{\mathbb{R}^{3N}}d^{3N}\mathbf{q}\,\rho(q,t)\,\bar{\theta}_{joint}^{lab}(q(t),t) & =\frac{1}{\hbar}\int_{\mathbb{R}^{3N}}d^{3N}\mathbf{q}\,\rho\,\left[\int_{t_{I}}^{t_{F}}E_{joint}dt-\sum_{i=1}^{N}\int_{q_{iI}}^{q_{iF}}m_{i}\mathbf{v}_{i}\cdot d\mathbf{q}_{i}(t)\right]+\sum_{i=1}^{N}\int_{\mathbb{R}^{3N}}d^{3N}\mathbf{q}\,\rho\,\phi_{i}\\
 & =\frac{1}{\hbar}\left(\int_{t_{I}}^{t_{F}}<E_{joint}>dt-\sum_{i=1}^{N}\int_{q_{iI}}^{q_{iF}}<\mathbf{p}_{i}\cdot d\mathbf{q}_{i}(t)>\right)+\sum_{i=1}^{N}\phi_{i},\\
 & =\int_{\mathbb{R}^{3N}}d^{3N}\mathbf{q}\,\rho\int_{t_{I}}^{t_{F}}\left\{ \sum_{i=1}^{N}\left[m_{i}c^{2}+\frac{1}{2}m_{i}v_{i}^{2}+\frac{1}{2}m_{i}u_{i}^{2}\right]-V_{g}^{int}-V_{c}^{int}\right\} dt+\sum_{i=1}^{N}\phi_{i},
\end{aligned}
\end{equation}
which follows from the time-symmetric mean Legendre transformation
\begin{equation}
L\coloneqq\sum_{i=1}^{N}\frac{1}{2}\left[(m\mathbf{b}_{i})\cdot\mathbf{b}_{i}+(m\mathbf{b}_{i*})\cdot\mathbf{b}_{i*}\right]-\frac{1}{2}\left(E_{joint+}+E_{joint-}\right)=\sum_{i=1}^{N}(m_{i}\mathbf{v}_{i})\cdot\mathbf{v}_{i}+(m_{i}\mathbf{u}_{i})\cdot\mathbf{u}_{i}-E_{joint},
\end{equation}
and using $\bar{\theta}_{joint}^{lab}=-\frac{1}{\hbar}S$. Applying 

\begin{equation}
J(q,t)=extremal,
\end{equation}
we have \cite{Derakhshani2016b}

\begin{equation}
\sum_{i=1}^{N}\frac{m_{i}}{2}\left[D_{*}D+DD_{*}\right]\mathbf{q}_{i}(t)=-\sum_{i=1}^{N}\nabla_{i}\frac{\left[m_{i}\Phi_{g}(\mathbf{q}_{i},\mathbf{q}_{j})+e_{i}\Phi_{c}(\mathbf{q}_{i},\mathbf{q}_{j})\right]}{2}|_{\mathbf{q}_{j}=\mathbf{q}_{j}(t)},
\end{equation}
for $i,j=1,...,N$ and $i\neq j$. And from the independent $\delta\mathbf{q}_{i}(t)$,
we have the individual equations of motion

\begin{equation}
m_{i}\mathbf{a}_{i}(q(t),t)\coloneqq\frac{m_{i}}{2}\left[D_{*}D+DD_{*}\right]\mathbf{q}_{i}(t)=-\nabla_{i}\frac{\left[m_{i}\Phi_{g}(\mathbf{q}_{i},\mathbf{q}_{j})+e_{i}\Phi_{c}(\mathbf{q}_{i},\mathbf{q}_{j})\right]}{2}|_{\mathbf{q}_{j}=\mathbf{q}_{j}(t)}.
\end{equation}
Applying the mean derivatives, using that $\mathbf{b}_{i}=\mathbf{v}_{i}+\mathbf{u}_{i}$,
$\mathbf{b}_{i*}=\mathbf{v}_{i}-\mathbf{u}_{i}$, and replacing $q(t)$
by $q$ on both sides, (58) becomes

\begin{equation}
\begin{aligned} & \sum_{i=1}^{N}m_{i}\left[\partial_{t}\mathbf{v}_{i}+\mathbf{v}_{i}\cdot\nabla_{i}\mathbf{v}_{i}-\mathbf{u}_{i}\cdot\nabla_{i}\mathbf{u}_{i}-\frac{\hbar}{2m_{i}}\nabla_{i}^{2}\mathbf{u}_{i}\right]\\
 & =-\sum_{i=1}^{N}\nabla_{i}\frac{\left[m_{i}\Phi_{g}(\mathbf{q}_{i},\mathbf{q}_{j})+e_{i}\Phi_{c}(\mathbf{q}_{i},\mathbf{q}_{j})\right]}{2}.
\end{aligned}
\end{equation}
Identifying 

\begin{equation}
\mathbf{p}_{i}=-\hbar\nabla_{i}\bar{\theta}_{joint}^{lab}=\nabla_{i}S,
\end{equation}
using (16-17) in (60), integrating both sides, and setting the arbitrary
integration constants equal to the particle rest energies, we then
get

\begin{equation}
\begin{aligned}-\partial_{t}S(q,t) & =\sum_{i=1}^{N}m_{i}c^{2}+\sum_{i=1}^{N}\frac{\left[\nabla_{i}S(q,t)\right]^{2}}{2m_{i}}\\
 & +\sum_{i=1}^{N}\frac{\left[m_{i}\Phi_{g}(\mathbf{q}_{i},\mathbf{q}_{j})+e_{i}\Phi_{c}(\mathbf{q}_{i},\mathbf{q}_{j})\right]}{2}-\sum_{i=1}^{N}\frac{\hbar^{2}}{2m_{i}}\frac{\nabla_{i}^{2}\sqrt{\rho(q,t)}}{\sqrt{\rho(q,t)}},
\end{aligned}
\end{equation}
with general solution 

\begin{equation}
S=\left(\sum_{i=1}^{N}\int\mathbf{p}_{i}\cdot d\mathbf{q}_{i}-\int Edt\right)-\sum_{i=1}^{N}\hbar\phi_{i}.
\end{equation}

Recall we made the plausible assumption that the presence of classical
external potentials doesn't alter the harmonic nature of the mean
\emph{zbw} oscillations of the particles. Hence, each \emph{zbw} particle
has a well-defined mean phase along its time-symmetric mean trajectory.
Accordingly, when $\Phi_{g,c}$ is not negligible, the joint phase
is a well-defined function of the time-symmetric mean trajectories
of\emph{ }all the \emph{zbw} particles. 

So for a closed loop \emph{L} along which each \emph{zbw} particle
can be physically or virtually displaced, it follows that 

\begin{equation}
\sum_{i=1}^{N}\oint_{L}\delta_{i}S=\sum_{i=1}^{N}\oint_{L}\left[\mathbf{p}_{i}\cdot\delta\mathbf{q}_{i}(t)-E\delta t\right]=nh,
\end{equation}
and 

\begin{equation}
\sum_{i=1}^{N}\oint_{L}\delta_{i}S=\sum_{i=1}^{N}\oint_{L}\mathbf{p}_{i}\cdot\delta\mathbf{q}_{i}(t)=\sum_{i=1}^{N}\oint_{L}\mathbf{\nabla}_{i}S|_{\mathbf{q}_{j}=\mathbf{q}_{j}(t)}\cdot\delta\mathbf{q}_{i}(t)=nh,
\end{equation}
the latter for a closed loop $L$ with $\delta t=0$. For the joint
mean phase field $S(q,t)$, applying the same physical reasoning to
each member of the \emph{i}-th ensemble yields

\begin{equation}
\sum_{i=1}^{N}\oint_{L}d_{i}S=\sum_{i=1}^{N}\oint_{L}\mathbf{p}_{i}\cdot d\mathbf{q}_{i}=\sum_{i=1}^{N}\oint_{L}\mathbf{\nabla}_{i}S\cdot d\mathbf{q}_{i}=nh.
\end{equation}
Clearly (66) implies phase-field quantization for each ensemble, upon
keeping all but the \emph{i}-th coordinate fixed and performing the
closed-loop integration. Applying the Madelung transformation to the
combination of (66), (62), and (52), we can construct the \emph{N}-particle
Schr{\"o}dinger equation for classically interacting \emph{zbw} particles
in the presence of external fields:

\begin{equation}
i\hbar\frac{\partial\psi(q,t)}{\partial t}=\sum_{i=1}^{N}\left[-\frac{\hbar^{2}}{2m_{i}}\nabla_{i}^{2}+m_{i}c^{2}+\frac{m_{i}\hat{\Phi}_{g}(\mathbf{\hat{q}}_{i},\mathbf{\hat{q}}_{j})}{2}+\frac{e_{i}\hat{\Phi}_{c}(\mathbf{\hat{q}}_{i},\mathbf{\hat{q}}_{j})}{2}\right]\psi(q,t),
\end{equation}
where $\psi(q,t)=\sqrt{\rho(q,t)}e^{iS(q,t)/\hbar}$ is single-valued
via (66).

Note the inclusion of hats on the interaction potentials and their
coordinates, in contrast to the quantum Hamilton-Jacobi (QHJ) equation
(62). As shown by Holland \cite{HollandBook1993} and Oriols \& Mompart
\cite{OriolsMompart2012}, there exists a correspondence between quantum
operators in the Schr{\"o}dinger equation, and c-number variables in
the QHJ equation. For example, the quantum expectation value of the
position operator corresponds to the ensemble averaged position coordinate
via $\left\langle \psi\right|\hat{\mathbf{q}}\left|\psi\right\rangle =\int_{\mathbb{R}^{3N}}d^{3N}\mathbf{q}\,\psi^{*}\,\mathbf{\hat{q}\,}\psi=\int_{\mathbb{R}^{3N}}d^{3N}\mathbf{q}\,\rho\,\mathbf{q}=<\mathbf{q}>$.
For another example, the quantum expectation value of the Hamiltonian
operator is equivalent to the ensemble average of the total energy
in the QHJ equation:

\begin{equation}
\begin{aligned}\left\langle \psi\right|\hat{H}\left|\psi\right\rangle  & =\int_{\mathbb{R}^{3N}}d^{3N}\mathbf{q}\,\psi^{*}(q,t)\left(\sum_{i=1}^{N}\left[-\frac{\hbar^{2}}{2m_{i}}\nabla_{i}^{2}+m_{i}c^{2}+\frac{m_{i}\hat{\Phi}_{g}(\mathbf{\hat{q}}_{i},\mathbf{\hat{q}}_{j})}{2}+\frac{e_{i}\hat{\Phi}_{c}(\mathbf{\hat{q}}_{i},\mathbf{\hat{q}}_{j})}{2}\right]\right)\psi(q,t)\\
 & =\int_{\mathbb{R}^{3N}}d^{3N}\mathbf{q}\,\rho(q,t)\left(\sum_{i=1}^{N}\left[m_{i}c^{2}+\frac{\left[\nabla_{i}S(q,t)\right]^{2}}{2m_{i}}+\frac{m_{i}\Phi_{g}(\mathbf{q}_{i},\mathbf{q}_{j})}{2}+\frac{e_{i}\Phi_{c}(\mathbf{q}_{i},\mathbf{q}_{j})}{2}-\frac{\hbar^{2}}{2m_{i}}\frac{\nabla_{i}^{2}\sqrt{\rho(q,t)}}{\sqrt{\rho(q,t)}}\right]\right)=\,<H>.
\end{aligned}
\end{equation}

So the classical potentials are, in effect, `quantized' at the level
of the Schr{\"o}dinger equation, insofar as they depend on operator-valued
position coordinates and satisfy the Poisson equations \footnote{The use of delta functions in the definitions of the mass and charge
densities is justified because we are using the point-like approximation
for interactions between the particles. In actuality, the mass and
charge densities should presumably depend on some form-factor $f(|\mathbf{x}-\mathbf{q}_{i}|)$
which distributes the mass or charge of the particle on its Compton
length-scale $\lambda_{c}$. Additionally, in scattering events where
$|\mathbf{q}_{i}(t)-\mathbf{q}_{j}(t)|\sim\lambda_{c}$ the point-like
approximation will no longer hold and it will become necessary to
include this form-factor in calculating the interactions. The precise
expression for this form-factor will depend on the specific physical
model used for the \emph{zbw} particle, which at present we do not
have (although see section 5 of \cite{Derakhshani2016b} for a discussion
of possibilities). Nevertheless, as we are only concerned here with
the non-relativistic regime, the point-like approximation will suffice.}
\begin{equation}
\nabla^{2}\hat{\Phi}_{g}=4\pi\sum_{i=1}^{N}m_{i}\delta^{3}\left(\mathbf{q}-\hat{\mathbf{q}}_{i}\right),
\end{equation}
 
\begin{equation}
\nabla^{2}\hat{\Phi}_{c}=-4\pi\sum_{i=1}^{N}e_{i}\delta^{3}\left(\mathbf{q}-\hat{\mathbf{q}}_{i}\right).
\end{equation}
Accordingly, the equation set (67-70) gives a statistical mechanical
description of $N$ \emph{zbw} particles undergoing Nelsonian diffusions,
while interacting both gravitationally and electrostatically through
the classical potentials (39-40). 

Equations (67-70) correspond to the standard quantum mechanical equations
for $N$ particles interacting gravitationally or electrostatically
in the Newtonian regime \cite{BardosGolseMauser2000,BardosErdosGolseMauserYau2002,Golse2003,AnastopoulosHu2014},
and that the standard quantum mechanical equations are the Newtonian
limits of the standard theories of perturbatively quantized gravity
and perturbative quantum electrodynamics \cite{AnastopoulosHu2014,BahramiBassi2014a}.
But because we derived (67-70) within the ZSM framework, we can go
further than the standard quantum description. That is, we can use
solutions of (67), or the equivalent solutions of the Madelung equations
(52) and (62) under the constraint (66), to deduce an ensemble of
possible trajectories for the actual (\emph{zbw}) particles.

In particular, it is readily shown that the \emph{i}-th mean acceleration

\begin{equation}
\begin{aligned}m_{i}\mathbf{a}_{i}(q(t),t) & =\frac{m_{i}}{2}\left[D_{*}D+DD_{*}\right]\mathbf{q}_{i}(t)=-\nabla_{i}\frac{\left[m_{i}\Phi_{g}(\mathbf{q}_{i},\mathbf{q}_{j})+e_{i}\Phi_{c}(\mathbf{q}_{i},\mathbf{q}_{j})\right]}{2}|_{\mathbf{q}_{j}=\mathbf{q}_{j}(t)}\\
 & \downarrow\\
m_{i}\frac{\mathrm{D}\mathbf{v}_{i}(q(t),t)}{\mathrm{D}t} & =\left[\partial_{t}\mathbf{p}_{i}+\mathbf{v}_{i}\cdot\nabla_{i}\mathbf{p}_{i}\right](q,t)|_{\mathbf{q}_{j}=\mathbf{q}_{j}(t)}=-\nabla_{i}\left[\frac{m_{i}\Phi_{g}(\mathbf{q}_{i},\mathbf{q}_{j})}{2}+\frac{e_{i}\Phi_{c}(\mathbf{q}_{i},\mathbf{q}_{j})}{2}-\frac{\hbar^{2}}{2m_{i}}\frac{\nabla_{i}^{2}\sqrt{\rho(q,t)}}{\sqrt{\rho(q,t)}}\right]|_{\mathbf{q}_{j}=\mathbf{q}_{j}(t)},
\end{aligned}
\end{equation}
where $\mathrm{D}/\mathrm{D}t$ is the convective derivative. Integrating
this last equation for different possible initial conditions $\mathbf{q}_{j}(0)$
allows us to construct an ensemble of mean trajectories, only one
of which is realized by the actual $i$-th\emph{ zbw} particle. We
can then find the gravitational and Coulomb potentials sourced by
the actual \emph{zbw} particles, along their mean trajectories, as
follows: 
\begin{equation}
\nabla^{2}\Phi_{g}^{m.t.}=4\pi\sum_{i=1}^{N}m_{i}\delta^{3}\left(\mathbf{q}-\mathbf{q}_{i}(t)\right),
\end{equation}
 
\begin{equation}
\nabla^{2}\Phi_{c}^{m.t.}=-4\pi\sum_{i=1}^{N}e_{i}\delta^{3}\left(\mathbf{q}-\mathbf{q}_{i}(t)\right),
\end{equation}
where the superscript ``$m.t.$'' refers to the interaction potentials
sourced by the mean trajectories of the actual \emph{zbw} particles. 

Actually, (72) doesn't contain all the terms that contribute to the
total mass-densities of the particles. The complete expression is
\begin{equation}
\nabla^{2}\Phi_{g}^{m.t.}=4\pi\sum_{i=1}^{N}\left[m_{i}+\frac{\left[\nabla_{i}S\left(\mathbf{q}_{i}(t),t\right)\right]^{2}}{2m_{i}c^{2}}-\frac{\hbar^{2}}{2mc^{2}}\frac{\nabla_{i}^{2}\sqrt{\rho\left(\mathbf{q}_{mi}(t),t\right)}}{\sqrt{\rho\left(\mathbf{q}_{mi}(t),t\right)}}\right]\delta^{3}\left(\mathbf{q}-\mathbf{q}_{i}(t)\right).
\end{equation}
But in the $v_{i}\ll c$ limit, the classical kinetic and quantum
kinetic \footnote{While the latter terms are referred to in the literature as ``quantum
potentials'' \cite{Broglie1930,Bohm1952I,Bohm1952II,BohmHiley1993,HollandBook1993,Duerr2009,OriolsMompart2012},
we prefer the term ``quantum kinetics'' \cite{Derakhshani2016a,Derakhshani2016b}
since, in stochastic mechanics, they arise from the kinetic energy
contributions of the osmotic velocities of the particles, as seen
from the left hand side of (60). } energy terms are negligible relative to the rest-energy terms, allowing
us to effectively neglect the contributions of the kinetic energies
to the total mass-energy density of the particle.

From the solutions of (67), we can also construct an ensemble of possible
stochastic trajectories for the $i$-th particle: 

\begin{equation}
d\mathbf{q}_{i}(t)=\left[\frac{\hbar}{m_{i}}\mathrm{Im}\frac{\nabla_{i}\psi(q,t)}{\psi(q,t)}+\frac{\hbar}{m_{i}}\mathrm{Re}\frac{\nabla_{i}\psi(q,t)}{\psi(q,t)}\right]|_{\mathbf{q}_{j}=\mathbf{q}_{j}(t)}dt+d\mathbf{W}_{i}(t),
\end{equation}

\begin{equation}
d\mathbf{q}_{i}(t)=\left[\frac{\hbar}{m_{i}}\mathrm{Im}\frac{\nabla_{i}\psi(q,t)}{\psi(q,t)}-\frac{\hbar}{m_{i}}\mathrm{Re}\frac{\nabla_{i}\psi(q,t)}{\psi(q,t)}\right]|_{\mathbf{q}_{j}=\mathbf{q}_{j}(t)}dt+d\mathbf{W}_{i*}(t).
\end{equation}
These stochastic trajectories can also be used in the definition of
the mass and charge densities, implying \emph{classically fluctuating}
mass and charge densities, hence \emph{classically fluctuating} gravitational
and Coulomb potentials satisfying the Poisson equations 
\begin{equation}
\nabla^{2}\Phi_{g}^{s.t.}=4\pi\sum_{i=1}^{N}m_{i}\delta^{3}\left(\mathbf{q}-\mathbf{q}_{i}(t)\right),
\end{equation}
 
\begin{equation}
\nabla^{2}\Phi_{c}^{s.t.}=-4\pi\sum_{i=1}^{N}e_{i}\delta^{3}\left(\mathbf{q}-\mathbf{q}_{i}(t)\right),
\end{equation}
where ``$s.t.$'' refers to the interaction potentials sourced by
the stochastic trajectories of the actual \emph{zbw} particles. 

Thus we see here that there are three `levels' of interaction potentials,
with $\Phi_{g,c}^{s.t.}$ being the most fundamental (in the sense
of being the potentials sourced by the actual, stochastic trajectories
of the actual \emph{zbw} particles), followed by $\Phi_{g,c}^{m.t.}$
(in the sense of being the potentials sourced by the mean trajectories
of the actual \emph{zbw} particles), and then $\Phi_{g,c}$ or $\hat{\Phi}_{g,c}$
(in the sense of being operator-valued potentials that reflect a statistical
ensemble of possible potentials sourced by the possible mean trajectories
of the actual \emph{zbw} particles). Indeed the operator-valued interaction
potentials $\hat{\Phi}_{g,c}$ have physical meaning inasmuch as 
\begin{equation}
\begin{aligned}\left\langle \psi\right|\hat{V}_{g}^{int}\left|\psi\right\rangle  & =\int_{\mathbb{R}^{3N}}d^{3N}\mathbf{q}\,\psi^{*}\hat{V}_{g}^{int}\,\psi=\int_{\mathbb{R}^{3N}}d^{3N}\mathbf{q}\,\rho\,V_{g}^{int}=<V_{g}^{int}>,\end{aligned}
\end{equation}
and

\begin{equation}
\begin{aligned}\left\langle \psi\right|\nabla^{2}\hat{\Phi}_{g}\left|\psi\right\rangle  & =4\pi\left\langle \psi\right|\sum_{i=1}^{N}m_{i}\delta^{3}\left(\mathbf{q}-\hat{\mathbf{q}}_{i}\right)\left|\psi\right\rangle \\
 & =4\pi\underset{i=1}{\overset{N}{\sum}}\int d^{3}\mathbf{q}'_{1}...d^{3}\mathbf{q}'_{N}|\psi(\mathbf{q}_{1}'...\mathbf{q}_{N}',t)|^{2}m_{i}\delta^{3}(\mathbf{q}-\mathbf{q}'_{i})\\
 & =4\pi\sum_{i=1}^{N}\int_{\mathbb{R}^{3N}}d^{3N}\mathbf{q}'\rho(q',t)\,m_{i}\delta^{3}\left(\mathbf{q}-\mathbf{q}_{i}(t)\right)|_{\mathbf{q}_{i}(t)=\mathbf{q}'_{i}}=<\nabla^{2}\Phi_{g}>,
\end{aligned}
\end{equation}
 and likewise for the Coulomb potentials.

Note the conceptual difference between the expected values of the
interaction potentials, equation (79), and the potentials obtained
from (72-73) (i.e., the potentials sourced by the mean trajectories
of the actual \emph{zbw} particles). The former are obtained from
averaging the interaction potentials over N statistical ensembles
of mean trajectories; the latter are obtained from taking the conditional
expectation at time \emph{t }of (75-76), considering the time-symmetrized
mean (current) velocity $\mathbf{v}=\left(1/2\right)\left(\mathbf{b}+\mathbf{b}_{*}\right)$,
and using the corresponding mean trajectories to source the potentials. 

It is interesting to compare (80) to the Poisson equation associated
with the \emph{N}-body Schr{\"o}dinger-Newton (SN) gravitational potential
\cite{Diosi1984,Adler2007,AnastopoulosHu2014,Derakhshani2014,BahramiBassi2014a,Giulini2014,DerakProbingGravCat2016}:
\begin{equation}
\nabla^{2}\Phi_{g}^{SN}=4\pi\underset{i=1}{\overset{N}{\sum}}\int d^{3}\mathbf{q}'_{1}...d^{3}\mathbf{q}'_{N}|\psi(\mathbf{q}_{1}'...\mathbf{q}_{N}',t)|^{2}m_{i}\delta^{3}(\mathbf{q}-\mathbf{q}'_{i}),
\end{equation}
In the SN equations, the solution of (81) describes the net interaction
potential sourced by \emph{N} matter density fields on space-time
(each field corresponding to an elementary `particle'), and this potential
feeds back into the Hamiltonian of the Schr{\"o}dinger equation to generate
a nonlinear Schr{\"o}dinger evolution. In ZSM-Newton, by contrast, the
solution of (80) describes the ensemble-average of the net interaction
potential sourced by the \emph{N} point-like \emph{zbw} particles,
and this potential \emph{does not} feed back into the (derived) Schr{\"o}dinger
Hamiltonian. Everything said here also holds for the Coulombic analogue
of (80) and its comparison to the \emph{N}-body Schr{\"o}dinger-Coulomb
equations \cite{BardosGolseMauser2000,BardosErdosGolseMauserYau2002,Golse2003,Derakhshani2014,AnastopoulosHu2014}.
(See subsection 5.1 for a more detailed comparison of ZSM-Newton/Coulomb
to N-body Schr{\"o}dinger-Newton/Coulomb.) 

Earlier we observed that the complete expression for the mass densities
of the \emph{zbw} particles is given by the rhs of (74). While we
also noted that the classical and quantum kinetic energy terms can
be neglected in the Newtonian regime, let us see what happens if we
do use the solution of (74) in the QHJ equation (62) and the Schr{\"o}dinger
equation (67). For maximum clarity, we restrict to the two-particle
case $q=\left\{ \mathbf{q}_{1},\mathbf{q}_{2}\right\} $ and drop
the Coulomb potentials and rest-energy terms: 

\begin{equation}
\begin{aligned}-\partial_{t}S(q,t) & =\sum_{i=1}^{2}\left(\frac{\left[\nabla_{i}S(q,t)\right]^{2}}{2m_{i}}-\frac{\hbar^{2}}{2m_{i}}\frac{\nabla_{i}^{2}\sqrt{\rho(q,t)}}{\sqrt{\rho(q,t)}}\right)-\frac{\left[m_{1}+\frac{T_{1}(q,t)}{c^{2}}+\frac{Q_{1}(q,t)}{c^{2}}\right]\left[m_{2}+\frac{T_{2}(q,t)}{c^{2}}+\frac{Q_{2}(q,t)}{c^{2}}\right]}{|\mathbf{q}_{1}-\mathbf{q}_{2}|}\\
 & =\sum_{i=1}^{2}\left(\frac{\left[\nabla_{i}S\right]^{2}}{2m_{i}}-\frac{\hbar^{2}}{2m_{i}}\frac{\nabla_{i}^{2}\sqrt{\rho}}{\sqrt{\rho}}\right)-\frac{\left[m_{1}m_{2}+m_{1}\frac{T_{2}}{c^{2}}+m_{1}\frac{Q_{2}}{c^{2}}+m_{2}\frac{T_{1}}{c^{2}}+m_{2}\frac{Q_{1}}{c^{2}}+\frac{T_{1}T_{2}}{c^{4}}+\frac{T_{1}Q_{2}}{c^{4}}+\frac{Q_{1}T_{2}}{c^{4}}+\frac{Q_{1}Q_{2}}{c^{4}}\right]}{|\mathbf{q}_{1}-\mathbf{q}_{2}|},
\end{aligned}
\end{equation}
where $T_{i}(q,t)\coloneqq\frac{\left[\nabla_{i}S(q,t)\right]^{2}}{2m_{i}}$
and $Q_{i}(q,t)\coloneqq-\frac{\hbar^{2}}{2m_{i}}\frac{\nabla_{i}^{2}\sqrt{\rho(q,t)}}{\sqrt{\rho(q,t)}}$.
We can see that the gravitational interaction energy between the two
particles depends on their classical kinetic and quantum kinetic energy
terms, along with their rest masses. Furthermore, using the Madelung
transformation to combine (82) with the continuity equation (52),
we obtain the \emph{nonlinear} two-particle Schr{\"o}dinger equation 

\begin{equation}
i\hbar\frac{\partial\psi(q,t)}{\partial t}=-\sum_{i=1}^{2}\frac{\hbar^{2}}{2m_{i}}\nabla_{i}^{2}\psi(q,t)-\left(\frac{\left[m_{1}m_{2}+\frac{m_{1}T_{2}}{c^{2}}+\frac{m_{1}Q_{2}}{c^{2}}+\frac{m_{2}T_{1}}{c^{2}}+\frac{m_{2}Q_{1}}{c^{2}}+\frac{T_{1}T_{2}}{c^{4}}+\frac{T_{1}Q_{2}}{c^{4}}+\frac{T_{2}Q_{1}}{c^{4}}+\frac{Q_{1}Q_{2}}{c^{4}}\right]}{|\mathbf{q}_{1}-\mathbf{q}_{2}|}\right)\psi(q,t),
\end{equation}
where 
\begin{equation}
\frac{m_{i}T_{j}}{c^{2}}=\frac{m_{i}}{2m_{j}c^{2}}\left[\nabla_{j}S\right]^{2}=\frac{\hbar^{2}m_{i}}{2c^{2}m_{j}^{2}}\left(\nabla_{j}\ln\psi\right)^{2},
\end{equation}
 
\begin{equation}
\frac{m_{i}Q_{j}}{c^{2}}=-\frac{\hbar^{2}m_{i}}{2c^{2}m_{j}^{2}}\frac{\nabla_{j}^{2}\sqrt{\rho}}{\sqrt{\rho}}=-\frac{\hbar^{2}m_{i}}{2c^{2}m_{j}^{2}}\frac{\nabla_{j}^{2}|\psi|}{|\psi|},
\end{equation}
\begin{equation}
\frac{T_{1}T_{2}}{c^{4}}=\frac{\left[\nabla_{1}S\right]^{2}\left[\nabla_{2}S\right]^{2}}{4m_{1}m_{2}c^{4}}=\frac{\hbar^{2}}{4c^{4}m_{1}m_{2}}\left(\nabla_{1}\ln\psi\right)^{2}\left(\nabla_{2}\ln\psi\right)^{2},
\end{equation}
\begin{equation}
\frac{T_{i}Q_{j}}{c^{4}}=-\frac{\hbar^{2}}{4m_{i}m_{j}c^{4}}\left[\nabla_{i}S\right]^{2}\frac{\nabla_{j}^{2}\sqrt{\rho}}{\sqrt{\rho}}=-\frac{\hbar^{2}}{4m_{i}m_{j}c^{4}}\left(\nabla_{i}\ln\psi\right)^{2}\frac{\nabla_{j}^{2}|\psi|}{|\psi|},
\end{equation}
\begin{equation}
\frac{Q_{1}Q_{2}}{c^{4}}=\frac{\hbar^{4}}{4m_{1}m_{2}c^{4}}\frac{\left(\nabla_{1}^{2}\sqrt{\rho}\right)\left(\nabla_{2}^{2}\sqrt{\rho}\right)}{\rho}=\frac{\hbar^{4}}{4m_{1}m_{2}c^{4}}\frac{\left(\nabla_{1}^{2}|\psi|\right)\left(\nabla_{2}^{2}|\psi|\right)}{|\psi|^{2}},
\end{equation}
for $i\neq j$ and $\psi=\sqrt{\rho}e^{iS/\hbar}$. 

Because of the nonlinearity of (83), the 3-space coordinates $\mathbf{q}_{1}$
and $\mathbf{q}_{2}$ in the Green's function of the gravitational
potential in (83) can no longer be interpreted as linear operators
(hence why we don't put hats on them) and $\psi$ no longer has a
consistent Born-rule interpretation \cite{Diosi2016}. (That $\psi$
of (83) has no consistent Born-rule interpretation means that Salcedo's
``statistical consistency problem'' for quantum-classical hybrid
theories \cite{Salcedo2012,Salcedo2016} is not applicable in the
present context, since Salcedo's problem assumes the validity of the
Born rule and standard quantum measurement postulates for hybrid theories.)
Nevertheless, $\rho=|\psi^{2}|$ is still (by definition!) the stochastic
mechanical position probability density for the two-particle system
and still evolves by the continuity equation (52). The important conceptual
distinction here is that the Born-rule interpretation of $|\psi|^{2}$
refers to the probability per unit volume of possible outcomes of
projective position measurements on the two-particle system, while
the stochastic mechanical definition of $|\psi|^{2}$ refers to the
probability per unit volume for the particles to \emph{be} at 3-space
positions $\{\mathbf{q}_{1},\mathbf{q}_{2}\}$ at time $t$ as a result
of their stochastic evolutions via (77-78). Thus, a break-down of
the Born-rule interpretation does not entail a break-down of the stochastic
mechanical meaning of $|\psi|^{2}$. 

Nonlinear Schr{\"o}dinger equations, together with entangled states,
are often said to imply superluminal signaling \cite{Gisin1989,Polchinski1991,Bacciagaluppi2012},
due to the well-known theorem of Gisin \cite{Gisin1989}. However,
as Bacciagaluppi has emphasized \cite{Bacciagaluppi2012}, superluminal
signaling only follows \emph{if} a theory with a nonlinear Schr{\"o}dinger
equation can also reproduce the usual phenomenology of wavefunction
collapse with Born-rule probabilities. Since said phenomenology does
not apply to solutions of (83), Gisin's theorem does not seem applicable
here. Of course, it may still be the case that the nonlinear Schr{\"o}dinger
equation (83) implies superluminal signaling, but determining this
depends on formulating a stochastic mechanical theory of measurement
consistent with (83). Since the nonlinearity of (83) makes naive application
of the standard stochastic mechanical theory of measurement \cite{Blanchard1986,Goldstein1987,Jibu1990,Blanchard1992,Peruzzi1996}
unreliable, it remains an open question what variant of the stochastic
mechanical theory of measurement is consistent with (83). However,
it is expected that such a variant will yield empirical predictions
in close agreement with the empirical predictions of the standard
stochastic mechanical theory of measurement applied to the linear
version of (83). The reason is that the nonlinear terms (84-88) are
ridiculously tiny in magnitude compared to the leading term proportional
to $m_{1}m_{2}$ in (83). So for all practical purposes, we can ignore
the nonlinear terms in modeling the Newtonian gravitational interaction
between the two \emph{zbw} particles, leaving us back to the linear
Schr{\"o}dinger equation (67).

What about the ether's gravitational contribution? The answer to this
will depend on the details of a proposed physical model of the ether,
which we currently do not have. Nevertheless, our phenomenological
hypotheses about the ether say that it is a medium in space-time with
superposed oscillations involving a countably infinite number of modes,
and that it continuously exchanges energy-momentum with the \emph{zbw}
particles. So it is reasonable to assume that there must be some stress-energy-momentum
associated with the ether. How this stress-energy-momentum gravitates
is an open question, but a couple possibilities can be noted: (i)
it doesn't gravitate at all, but rather the coupling of the ether
to massive \emph{zb}w particles somehow induces gravity on a Lorentzian
manifold, in analogy with Sakharov's `induced gravity' proposal \footnote{In Sakharov's approach, quantum vacuum fluctuations from matter fields
don't gravitate through their stress-energy-momentum tensor; rather,
one-loop vacuum fluctuations on a Lorentzian manifold (the latter
left to 'flap in the breeze') generate an effective action that contains
terms proportional to the Einstein-Hilbert action, the cosmological
constant, plus ``curvature-squared'' terms \cite{Visser2002}.} \cite{Visser2002}; (ii) it gravitates, but its overall contribution
to the total system energy density in the non-relativistic limit is
negligible compared to the rest-energy of a \emph{zbw} particle. In
our view, if the ether hypothesis of ZSM is correct, one of these
two possibilities must be correct, because all mass-energy quantities
experimentally measured in high energy scattering experiments and
nuclear binding/decay processes seem to come from three sources: (a)
the sum of the rest-masses of the particles, (b) the relativistic
kinetic energies of the particles, and (c) the mass-energy associated
with interactions between particles via the known fundamental forces.
Further supporting this view, we will show in a future paper that
a natural (semiclassical) general relativistic extension of ZSM suggests
a macroscopic model of the ether as a relativistic non-viscous fluid
that gravitates in the Einstein equations but gives a negligible contribution
to the total rest-energy of a system of \emph{zbw} particles in the
non-relativistic limit. Hence, for this paper, we shall continue with
neglecting the ether in gravitational effects (aside, of course, from
the ether's physical influence on the particles through their \emph{zbw}
oscillations and translational motions).

Finally, note that we have ignored the contribution of gravitational
and electrodynamical radiation reaction forces. In a separate paper,
we will show how these radiation reaction forces can be consistently
incorporated into ZSM-Newton/Coulomb through a stochastic generalization
of Galley's variational principle for nonconservative systems \cite{Galley2013}.

\section{Schr{\"o}dinger-Newton/Coulomb equations as mean-field theories}

We show in this section that ZSM-Newton/Coulomb recovers the `single-body'
Schr{\"o}dinger-Newton/Coulomb equations \cite{Diosi1984,Guzman2003,Adler2007,Carlip2008,Giulini2011,Bassi2013,Hu2014,Derakhshani2014,AnastopoulosHu2014,BahramiBassi2014a,Bera2015,DerakNewtLimStoGra2017,BardosGolseMauser2000,BardosErdosGolseMauserYau2002,Golse2003}
as mean-field approximations when the number of \emph{zbw} particles
is sufficiently large. For clarity, we separate out the gravitational
and Coulomb interactions. 

The main idea of a `mean-field' (or `large N') theory is to approximate
the evolution of many particles interacting (gravitationally and/or
electromagnetically), when N is large (i.e., $N\rightarrow\infty$)
and the interactions are weak (in the sense that the gravitational
coupling between particles scales as $1/N$) \cite{Golse2003}. So
for example, if a system of identical particles has the mean-field
phase-space density $f(\mathbf{q},\mathbf{p},t)$, the mean-field
approximation says that the force exerted on a typical particle in
the system by the N other particles is approximated by averaging -
with respect to the phase-space density - the force exerted on the
typical particle at its 3-space location, from each point in the phase
space. Mean-field theory can also be used to approximate the net (gravitational
and/or electrostatic) force from a cloud of many weakly interacting
identical particles, on an external (macroscopic or mesoscopic or
microscopic) body such as a force-measurement probe. 

It is instructive to first discuss the mean-field approximation scheme
for a classical system of weakly interacting particles. Let us consider
a slight variation on the example discussed by Golse in \cite{Golse2003},
namely, a system of N identical classical point particles, weakly
interacting gravitationally, with 6N-dimensional Hamiltonian

\begin{equation}
H\left(\mathbf{q}_{1}(t),...,\mathbf{q}_{N}(t);\mathbf{p}_{1}(t),...,\mathbf{p}_{N}(t)\right)=\sum_{i=1}^{N}\frac{p_{i}^{2}}{2m}+\frac{1}{N}V_{g}^{int},
\end{equation}
where $V_{g}^{int}(\mathbf{q}_{i}(t),\mathbf{q}_{j}(t))=\frac{1}{2}\sum_{i,j=1}^{N(j\neq i)}\frac{m^{2}}{|\mathbf{q}_{i}(t)-\mathbf{q}_{j}(t)|}$
and the $1/N$ factor is the `weak-coupling scaling' \footnote{Without the scaling, $V_{g}^{int}$ diverge much faster than the total
kinetic energy (a sum of N terms) as $N\rightarrow\infty$, since
the sum in $V_{g}^{int}$ is composed of $0.5\,N\left(N-1\right)$
terms. With the scaling, however, $N^{-1}V_{g}^{int}$ scales as N
in the $N\rightarrow\infty$ limit. Thus the weak coupling scaling
ensures that $V_{g}^{int}$ and the total kinetic energy scale in
the same way in the $N\rightarrow\infty$ limit.} . Physically, the Hamiltonian (89) describes a collisionless dilute
gas of gravitationally interacting non-relativistic particles, and
is a special case of the Hamiltonian considered by Golse \cite{Golse2003}
and Bardos et al. \cite{BardosGolseMauser2000,BardosErdosGolseMauserYau2002}
(they considered (89) for an arbitrary, symmetric, smooth interaction
potential). The dynamics for the point particles is generated by (89)
via Hamilton's equations $\dot{\mathbf{q}}_{i}(t)=m^{-1}\nabla_{\mathbf{p}_{i}}H$
and $\dot{\mathbf{p}}_{i}(t)=-\nabla_{\mathbf{q}_{i}}H$. Consider
now the empirical distribution for the N particles: $f_{N}(\mathbf{q},\mathbf{p},t)\coloneqq N^{-1}\sum_{i=1}^{N}\delta^{3}(\mathbf{q}-\mathbf{q}_{i}(t))\delta^{3}(\mathbf{p}-\mathbf{p}_{i}(t))$,
which satisfies (in the sense of distributions) the Vlasov equation
\begin{equation}
\partial_{t}f_{N}+\mathbf{p}\cdot\nabla_{\mathbf{q}}f_{N}+\nabla_{\mathbf{p}}\cdot\left[F_{N}\left(\mathbf{q},t\right)f_{N}\right]=\frac{1}{N^{2}}\sum_{i=1}^{N}\nabla_{\mathbf{p}}\cdot\left[\nabla_{\mathbf{q}}V\left(\mathbf{q}_{i},\mathbf{q}_{j}\right)\delta_{\mathbf{q}}\delta_{\mathbf{p}}\right],
\end{equation}
where
\begin{equation}
F_{N}\left(\mathbf{q},t\right)\coloneqq-\nabla_{\mathbf{q}}\int_{\mathbb{R}^{6}}\int_{\mathbb{R}^{6}}V\left(\mathbf{q},\mathbf{q}'\right)\,f_{N}\,d\mathbf{q}'d\mathbf{p}.
\end{equation}

Then, in the limit $N\rightarrow\infty$, the system described by
(89-91) is equivalent to a six-dimensional phase-space density $f(\mathbf{q},\mathbf{p},t)$
(representing the phase space density particles of mass $m$ located
at position $\mathbf{q}$ with momentum $\mathbf{p}$ at time $t$)
evolving by the `large N' Vlasov equation
\begin{equation}
\partial_{t}f(\mathbf{q},\mathbf{p},t)+\left\{ H^{m.f.}(\mathbf{q},\mathbf{p},t),f(\mathbf{q},\mathbf{p},t)\right\} =0,
\end{equation}
where the time-dependent ``mean-field'' Hamiltonian $H^{m.f.}(\mathbf{q},\mathbf{p},t)$
is given by

\begin{equation}
H^{m.f.}(\mathbf{q},\mathbf{p},t)=\frac{p^{2}}{2m}+\int_{\mathbb{R}^{3}}\int_{\mathbb{R}^{3}}m\Phi(\mathbf{q},\mathbf{q}')f(\mathbf{q}',\mathbf{p},t)d^{3}\mathbf{p}d^{3}\mathbf{q}'.
\end{equation}
The last term on the rhs of (93) is the ``mean-field'' potential
energy, i.e., the phase-space averaged potential energy of a typical
particle of mass $m$ at position $\mathbf{q}$ at time \emph{t}.
It can be rewritten as 

\begin{equation}
\int_{\mathbb{R}^{3}}m\Phi(\mathbf{q},\mathbf{q}')\left(\int_{\mathbb{R}^{3}}f(\mathbf{q}',\mathbf{p},t)d^{3}\mathbf{p}\right)d^{3}\mathbf{q}',
\end{equation}
which tells us it should be interpreted, more precisely, as the sum
of the elementary potentials created at position \emph{$\mathbf{q}$}
by one typical particle located at position \emph{$\mathbf{q}'$}
and distributed according to the 3-space particle number density

\begin{equation}
\rho(\mathbf{q}',t)\coloneqq\int_{\mathbb{R}^{3}}f(\mathbf{q}',\mathbf{p},t)d^{3}\mathbf{p},
\end{equation}
with normalization

\begin{equation}
\int_{\mathbb{R}^{3}}\rho(\mathbf{q},t)d^{3}\mathbf{q}=\int_{\mathbb{R}^{3}}\int_{\mathbb{R}^{3}}f(\mathbf{q},\mathbf{p},t)d^{3}\mathbf{q}d^{3}\mathbf{p}=N.
\end{equation}
 Since we are considering a system of identical particles interacting
gravitationally in the Newtonian approximation, the elementary potentials
are of the form 

\begin{equation}
\Phi(\mathbf{q},\mathbf{q}')=-\frac{m}{|\mathbf{q}-\mathbf{q}'|}.
\end{equation}
So with (94) and (95), we can rewrite (93) as 

\begin{equation}
H^{m.f.}(\mathbf{q},\mathbf{p},t)=\frac{p^{2}}{2m}-\int_{\mathbb{R}^{3}}\frac{m^{2}\rho(\mathbf{q}',t)}{|\mathbf{q}-\mathbf{q}'|}d^{3}\mathbf{q}',
\end{equation}
where $\rho(\mathbf{q},t)$ is the source in the Poisson equation

\begin{equation}
\nabla^{2}\Phi_{g}^{m.f.}=4\pi m\rho(\mathbf{q},t).
\end{equation}

Hence the mean-field Hamiltonian (93) describes, at time $t$, the
total energy of a typical particle with momentum $\mathbf{p}$ at
position $\mathbf{q}$ and with mean-field gravitational potential
energy $\int_{\mathbb{R}^{3}}m\Phi(\mathbf{q},\mathbf{q}')\rho(\mathbf{q}',t)d^{3}\mathbf{q}'$.
Correspondingly, the position-space number density (95) can be shown
to evolve by the continuity equation

\begin{equation}
\partial_{t}\rho(\mathbf{q},t)=-\nabla\cdot\left(\frac{\mathbf{p}(\mathbf{q},t)}{m}\rho(\mathbf{q},t)\right),
\end{equation}
upon projecting the Liouville equation for $f(\mathbf{q},\mathbf{p},t)$
into position space, where the mean momentum
\begin{equation}
\mathbf{p}(\mathbf{q},t)\coloneqq\int_{\mathbb{R}^{3}}\frac{\mathbf{p}f(\mathbf{q},\mathbf{p},t)d^{3}\mathbf{p}}{\rho(\mathbf{q},t)}.
\end{equation}

It is interesting to consider the special case when
\begin{equation}
f(\mathbf{q},\mathbf{p},t)=\rho(\mathbf{q},t)\delta^{3}\left[\mathbf{p}-\nabla S_{cl}(\mathbf{q},t)\right],
\end{equation}
where $S_{cl}(\mathbf{q},t)$ is a single-valued classical velocity
potential associated to a typical particle at position $\mathbf{q}$
at time $t$. We then have 
\begin{equation}
\mathbf{p}(\mathbf{q},t)=\int_{\mathbb{R}^{3}}\frac{\mathbf{p}f(\mathbf{q},\mathbf{p},t)d^{3}\mathbf{p}}{\rho(\mathbf{q},t)}=\nabla S_{cl}(\mathbf{q},t),
\end{equation}
and
\begin{equation}
H^{m.f.}(\mathbf{q},\nabla S_{cl},t)=\int_{\mathbb{R}^{3}}\frac{H^{m.f.}(\mathbf{q},\mathbf{p},t)f(\mathbf{q},\mathbf{p},t)d^{3}\mathbf{p}}{\rho(\mathbf{q},t)}=\frac{\left[\nabla S_{cl}(\mathbf{q},t)\right]^{2}}{2m}-\int_{\mathbb{R}^{3}}\frac{m^{2}\rho(\mathbf{q}',t)}{|\mathbf{q}-\mathbf{q}'|}d^{3}\mathbf{q}'.
\end{equation}
In this `Hamilton-Jacobi' case, (100) becomes

\begin{equation}
\partial_{t}\rho(\mathbf{q},t)=-\nabla\cdot\left(\frac{\nabla S_{cl}(\mathbf{q},t)}{m}\rho(\mathbf{q},t)\right),
\end{equation}
and (104) implies the Hamilton-Jacobi equation

\begin{equation}
H^{m.f.}(\mathbf{q},\nabla S_{cl},t)=-\partial_{t}S_{cl}(\mathbf{q},t)=\frac{\left[\nabla S_{cl}(\mathbf{q},t)\right]^{2}}{2m}-\int_{\mathbb{R}^{3}}\frac{m^{2}\rho(\mathbf{q}',t)}{|\mathbf{q}-\mathbf{q}'|}d^{3}\mathbf{q}'.
\end{equation}
Accordingly, the Madelung transformation on (105-106) yields the nonlinear
Schr{\"o}dinger equation 
\begin{equation}
i\hbar\partial_{t}\chi_{cl}(\mathbf{q},t)=\left(-\frac{\hbar^{2}}{2m}\nabla^{2}-\int d^{3}\mathbf{q}'\frac{m^{2}|\chi_{cl}(\mathbf{q}',t)|^{2}}{|\mathbf{q}-\mathbf{q}'|}+\frac{\hbar^{2}}{2m}\frac{\nabla^{2}\sqrt{|\chi_{cl}|}}{\sqrt{|\chi_{cl}|}}\right)\chi_{cl}(\mathbf{q},t),
\end{equation}
with corresponding Poisson equation 

\begin{equation}
\nabla^{2}\Phi_{g}^{m.f.}=4\pi m|\chi_{cl}(\mathbf{q},t)|^{2}.
\end{equation}
Here, $\chi_{cl}(\mathbf{q},t)=\sqrt{\rho(\mathbf{q},t)}e^{iS_{cl}(\mathbf{q},t)/\hbar}$
is the classical mean-field `wavefunction', a collective variable
describing the evolution of a large number of identical particles
that weakly interact gravitationally. Note that the set (107-108)
looks formally just like the single-body SN equations, but with the
addition of an opposite-signed quantum kinetic defined in terms of
the classical mean-field wavefunction. Likewise, if we had started
with the description of N identical charged particles weakly interacting
electrostatically, with Hamiltonian (89) under the replacement $V_{g}^{int}\rightarrow V_{c}^{int}$,
then by taking the large N limit and considering the Hamilton-Jacobi
case, we would obtain a nonlinear Schr{\"o}dinger-Coulomb-like system
identical to (107-108), with the charge $-e$ replacing the mass $m$.

We shall now develop a similar mean-field approximation scheme for
ZSM-Newton.

To model a dilute `gas' of N identical ZSM particles interacting weakly
through Newtonian gravitational forces, we introduce the N-particle
quantum Hamilton-Jacobi equation (for simplicity we drop the rest-energy
terms) with weak-coupling scaling:

\begin{equation}
\begin{aligned}-\partial_{t}S(q,t)|_{\mathbf{q}_{j}=\mathbf{q}_{j}(t)} & =\sum_{i=1}^{N}\frac{\left[\nabla_{i}S(q,t)\right]^{2}}{2m}|_{\mathbf{q}_{j}=\mathbf{q}_{j}(t)}+\frac{1}{N}V_{g}^{int}(\mathbf{q}_{i}(t),\mathbf{q}_{j}(t))-\sum_{i=1}^{N}\frac{\hbar^{2}}{2m}\frac{\nabla_{i}^{2}\sqrt{\rho(q,t)}}{\sqrt{\rho(q,t)}}|_{\mathbf{q}_{j}=\mathbf{q}_{j}(t)},\end{aligned}
\end{equation}
where $S$ satisfies 

\begin{equation}
\sum_{i=1}^{N}\oint_{L}\mathbf{\nabla}_{i}S|_{\mathbf{q}_{j}=\mathbf{q}_{j}(t)}\cdot\delta\mathbf{q}_{i}(t)=nh,
\end{equation}
for a closed loop $L$ with $\delta t=0$. 

Now, it is well-known in classical mechanics \cite{Jehle1953,Bohm2002,JoseSaletan}
that when harmonic oscillators of the same natural frequency are nonlinearly
coupled, they eventually synchronize and oscillate in phase with each
other. (The relative phase does oscillate, but in the long run those
oscillations average out to zero.) Since the \emph{zbw} particles
are essentially harmonic oscillators of identical natural frequencies
and are nonlinearly coupled via $V_{g}^{int}$, it is reasonable to
expect that, after some time, their oscillations eventually come into
phase with each other. When this `phase-locking' occurs between the
\emph{zbw} particles, we can plausibly make the ansatz that

\begin{equation}
S(q,0)=\sum_{i=1}^{N}S(\mathbf{q}_{i},0),
\end{equation}
where all the $S(\mathbf{q}_{i},0)$ are identical. 

Furthermore, since the N-particle continuity equation

\begin{equation}
\frac{\partial\rho({\normalcolor q},t)}{\partial t}=-\sum_{i=1}^{N}\nabla_{i}\cdot\left[\frac{\nabla_{i}S(q,t)}{m}\rho(q,t)\right],
\end{equation}
has the general solution
\begin{equation}
\rho(q,t)=e^{2R/\hbar}=\rho_{0}(q_{0})exp[-\int_{0}^{t}\left(\sum_{i}^{N}\nabla_{i}\cdot\frac{\nabla_{i}S}{m}\right)dt',
\end{equation}
the initial N-particle osmotic potential takes the form

\begin{equation}
R(q,t)=R_{0}(q_{0})-(\hbar/2)\int_{0}^{t}\left(\sum_{i=1}^{N}\nabla_{i}\cdot\frac{\nabla_{i}S}{m}\right)dt'.
\end{equation}
So it is also plausible to make the ansatz

\begin{equation}
R(q,0)=\sum_{i=1}^{N}R(\mathbf{q}_{i},0),
\end{equation}
where all the $R(\mathbf{q}_{i},0)$ are identical, which implies
that the initial N-particle probability density factorizes into a
product of identical single-particle densities:

\begin{equation}
\rho(q,0)=\prod_{i=1}^{N}\rho(\mathbf{q}_{i},0).
\end{equation}
From (116) it follows that (112) factorizes into N single-particle
continuity equations at $t=0$. Physically speaking, we can interpret
(115-116) as corresponding to the assumptions that, at $t=0$, the
way that the particle-ether coupling happens, in the local neighborhood
of each \emph{zbw} particle, is identical for all \emph{zbw} particles
(hence identical osmotic potentials sourced by the ether regions in
the local neighborhood of each \emph{zbw} particle), and that the
particles are interacting so weakly through $V_{g}^{int}$ and the
ether that they can be considered (effectively) physically independent
of one another.

Now, it is physically plausible to conjecture that, in the limit $N\rightarrow\infty$,
\footnote{Although we will not give a rigorous mathematical proof of this conjecture,
we will see later in this section that the conjecture is corroborated
by another large N argument that does have a rigorous mathematical
justification. Specifically, the large N limit prescription that leads
from the quantum N-body problem to the mean-field Schr{\"o}dinger-Poisson
equation that approximates a system of N quantum particles weakly
interacting by $1/r$ (e.g., Newtonian or Coulomb) potentials \cite{BardosGolseMauser2000,BardosErdosGolseMauserYau2002,Golse2003,AnastopoulosHu2014,BahramiBassi2014a,DerakNewtLimStoGra2017}. } the generation of correlations between the motions of the particles
get suppressed (because of the weak-coupling scaling) so that time-evolution
by (112) yields

\begin{equation}
\rho(q,t)=\prod_{i=1}^{N}\rho(\mathbf{q}_{i},t),
\end{equation}
and time-evolution by (109) yields

\begin{equation}
S(q,t)=\sum_{i=1}^{N}S(\mathbf{q}_{i},t),
\end{equation}
where $\rho(\mathbf{q},t)$ satisfies

\begin{equation}
\partial_{t}\rho(\mathbf{q},t)=-\nabla\cdot\left(\frac{\nabla S(\mathbf{q},t)}{m}\rho(\mathbf{q},t)\right),
\end{equation}
and $S(\mathbf{q},t)$ satisfies
\begin{equation}
-\partial_{t}S(\mathbf{q},t)=\frac{\left[\nabla S(\mathbf{q},t)\right]^{2}}{2m}+\int_{\mathbb{R}^{3}}m\Phi(\mathbf{q},\mathbf{q}')\rho(\mathbf{q}',t)d^{3}\mathbf{q}'-\frac{\hbar^{2}}{2m}\frac{\nabla^{2}\sqrt{\rho(\mathbf{q},t)}}{\sqrt{\rho(\mathbf{q},t)}},
\end{equation}
along with

\begin{equation}
\oint_{L}\mathbf{\nabla}S\cdot d\mathbf{q}=nh.
\end{equation}

Although $S(\mathbf{q},t)$ and $\rho(\mathbf{q},t)$ look formally
like single-particle variables, they are, in fact, collective variables
in a mean-field description of the exact many-body description given
by (109-110) with (111) and (115). In particular, $\rho(\mathbf{q},t)$
has the physical meaning of the density of \emph{zbw} particles of
mass $m$ occupying position $\mathbf{q}$ at time $t$. Similarly,
$S(\mathbf{q},t)$ is the \emph{zbw} phase of a typical \emph{zbw}
particle at $\mathbf{q}$ at time \emph{t}. Accordingly, the last
term on the right side of (120) is the quantum kinetic energy of the
typical \emph{zbw} particle at $\mathbf{q}$ at $t$, and
\begin{equation}
V_{g}^{m.f.}(\mathbf{q},t)=m\Phi_{g}^{m.f.}(\mathbf{q},t)=\int_{\mathbb{R}^{3}}m\Phi(\mathbf{q},\mathbf{q}')\rho(\mathbf{q}',t)d^{3}\mathbf{q}'
\end{equation}
 is the mean-field gravitational potential energy of the typical \emph{zbw}
particle at $\mathbf{q}$ at $t$, where $\Phi$ is the elementary
potential given by (97) and $\Phi_{g}^{m.f.}$ satisfies the Poisson
equation 

\begin{equation}
\nabla^{2}\Phi_{g}^{m.f.}=4\pi m\rho(\mathbf{q},t).
\end{equation}

It is worth observing that (119) can also be viewed as the position-space
projection of the modified Vlasov equation 
\begin{equation}
\partial_{t}f(\mathbf{q},\mathbf{p},t)+\frac{\mathbf{p}}{m}\cdot\nabla_{\mathbf{q}}f(\mathbf{q},\mathbf{p},t)+\mathbf{F}(\mathbf{q},t)\cdot\nabla_{\mathbf{p}}f(\mathbf{q},\mathbf{p},t)=0,
\end{equation}
where the initial phase-space density is defined by $f_{0}(\mathbf{q},\mathbf{p})\coloneqq\rho_{0}(\mathbf{q})\delta^{3}\left[\mathbf{p}-\nabla S_{0}(\mathbf{q})\right]$
and 
\begin{equation}
\begin{array}{c}
f_{0}(\mathbf{q},\mathbf{p})\coloneqq\rho_{0}(\mathbf{q})\delta^{3}\left[\mathbf{p}-\nabla S_{0}(\mathbf{q})\right]\\
\Downarrow\\
f(\mathbf{q},\mathbf{p},t)=\rho(\mathbf{q},t)\delta^{3}\left[\mathbf{p}-\nabla S(\mathbf{q},t)\right],
\end{array}
\end{equation}
due time-evolution by (119), along with normalization

\begin{equation}
\int_{\mathbb{R}^{3}}\rho(\mathbf{q},t)d^{3}\mathbf{q}=\int_{\mathbb{R}^{3}}\int_{\mathbb{R}^{3}}f(\mathbf{q},\mathbf{p},t)d^{3}\mathbf{q}d^{3}\mathbf{p}=N.
\end{equation}
From (125) it follows that the position-space projection of a typical
\emph{zbw} particle's 3-momentum $\mathbf{p}$ at position $\mathbf{q}$
yields 
\begin{equation}
\mathbf{p}(\mathbf{q},t)=\int_{\mathbb{R}^{3}}\frac{\mathbf{p}f(\mathbf{q},\mathbf{p},t)d^{3}\mathbf{p}}{\rho(\mathbf{q},t)}=\nabla S(\mathbf{q},t)
\end{equation}
for all times, where $\rho=\int_{\mathbb{R}^{3}}f\,d^{3}\mathbf{p}$.
The force term in (124) is 
\begin{equation}
\begin{split}\mathbf{F}(\mathbf{q},t) & \coloneqq-\nabla_{\mathbf{q}}\left[\int_{\mathbb{R}^{3}}\int_{\mathbb{R}^{3}}m\Phi(\mathbf{q},\mathbf{q}')f(\mathbf{q}',\mathbf{p},t)d^{3}\mathbf{p}d^{3}\mathbf{q}-\frac{\hbar^{2}}{2m}\frac{\nabla^{2}\sqrt{\rho(\mathbf{q},t)}}{\sqrt{\rho(\mathbf{q},t)}}\right]\\
 & =-\nabla_{\mathbf{q}}\left[\int_{\mathbb{R}^{3}}m\Phi(\mathbf{q},\mathbf{q}')\rho(\mathbf{q}',t)d^{3}\mathbf{q}-\frac{\hbar^{2}}{2m}\frac{\nabla^{2}\sqrt{\rho(\mathbf{q},t)}}{\sqrt{\rho(\mathbf{q},t)}}\right],
\end{split}
\end{equation}
and has the physical interpretation of the net force on a typical
\emph{zbw} particle at $\mathbf{q}$ at $t$, due to spatial gradients
of the mean-field gravitational potential energy \emph{and} quantum
kinetic energy of the typical \emph{zbw} particle at $\mathbf{q}$
at $t$. Correspondingly, it can be readily confirmed that the momentum-space
projection of (124), in conjunction with $f(\mathbf{q},\mathbf{p},t)=\rho(\mathbf{q},t)\delta^{3}\left[\mathbf{p}-\nabla S(\mathbf{q},t)\right]$,
yields \footnote{It is readily confirmed that the pressure tensor arising from the
momentum-space projection of (124) vanishes, because of the delta
function distribution in momentum in the definition of $f$. } 
\begin{equation}
\begin{split}\partial_{t}\mathbf{p}(\mathbf{q},t)+\mathbf{v}(\mathbf{q},t)\cdot\nabla\mathbf{p}(\mathbf{q},t) & =-\nabla_{\mathbf{q}}\left[\int_{\mathbb{R}^{3}}m\Phi(\mathbf{q},\mathbf{q}')\rho(\mathbf{q}',t)d^{3}\mathbf{q}-\frac{\hbar^{2}}{2m}\frac{\nabla^{2}\sqrt{\rho(\mathbf{q},t)}}{\sqrt{\rho(\mathbf{q},t)}}\right].\end{split}
\end{equation}

Now, applying the Madelung transformation to (119-121) yields the
mean-field nonlinear Schr{\"o}dinger equation 
\begin{equation}
i\hbar\partial_{t}\chi(\mathbf{q},t)=\left(-\frac{\hbar^{2}}{2m}\nabla^{2}-\int d^{3}\mathbf{q}'\frac{m^{2}|\chi(\mathbf{q}',t)|^{2}}{|\mathbf{q}-\mathbf{q}'|}\right)\chi(\mathbf{q},t),
\end{equation}
with corresponding Poisson equation 

\begin{equation}
\nabla^{2}\Phi_{g}^{m.f.}=4\pi m|\chi(\mathbf{q},t)|^{2},
\end{equation}
where $\chi(\mathbf{q},t)=\sqrt{\rho(\mathbf{q},t)}e^{iS(\mathbf{q},t)/\hbar}$.
Here, the mean-field wavefunction is, like the classical mean-field
wavefunction, a collective variable describing the evolution of a
large number of identical \emph{zbw} particles that weakly interact
gravitationally. We note that, this time, the set (130-131) formally
looks \emph{exactly} like the single-body SN equations, but with the
very different physical meaning as a mean-field approximation in the
sense just explained. Similarly, if we had started with the description
of N identical charged \emph{zbw} particles interacting electrostatically,
with QHJ equation (109) under the replacement $V_{g}^{int}\rightarrow V_{c}^{int}$,
then by taking the large N limit as prescribed above, we would get
a nonlinear Schr{\"o}dinger-Coulomb system identical to (130-131) with
$-e$ replacing $m$. 

Note that when the quantum kinetic and its first $\nabla_{\mathbf{q}}$
are negligible relative to the mean-field gravitational potential
energy and mean gravitational force, (130) effectively turns into
the classical nonlinear Schr{\"o}dinger equation (107), since (120) effectively
becomes (106). This observation seems to suggest a `quantum-classical'
correspondence between the Hamilton-Jacobi case of the classical Vlasov-Poisson
mean-field theory for a collisionless gas or plasma of non-relativistic
interacting particles, and the mean-field approximation for N-particle
ZSM-Newton/Coulomb. However, such a correspondence is only formal;
we will later see that the reliability of (130-131) as a mean-field
approximation breaks down for macroscopic superposition states.

To confirm the validity of our mean-field approximation proposal for
ZSM-Newton/Coulomb, let us reconsider the dilute gas of N identical
ZSM particles interacting through Newtonian gravitational forces,
but starting our description from the Schr{\"o}dinger equation (67) (minus
the rest-energy terms and the Coulomb potential) with weak-coupling
scaling:

\begin{equation}
i\hbar\frac{\partial\psi(q,t)}{\partial t}=\sum_{i=1}^{N}\left[-\frac{\hbar^{2}}{2m}+\frac{1}{N}\frac{m\hat{\Phi}_{g}(\mathbf{\hat{q}}_{i},\mathbf{\hat{q}}_{j})}{2}\right]\psi(q,t),
\end{equation}
where 
\begin{equation}
\nabla^{2}\hat{\Phi}_{g}=4\pi\sum_{i=1}^{N}m\delta^{3}\left(\mathbf{q}-\hat{\mathbf{q}}_{i}\right)
\end{equation}
and
\begin{equation}
\int_{\mathbb{R}^{3N}}|\psi(q,0)|^{2}d^{3N}\mathbf{q}=1.
\end{equation}
Supposing all the particles are in the same single-particle pure state
$\chi(\mathbf{q})$ at $t=0$, we can make the ``Hartree ansatz''

\begin{equation}
\psi(q,0)=\prod_{i=1}^{N}\chi(\mathbf{q}_{i},0),
\end{equation}
where the $\chi(\mathbf{q}_{i},0)$ are identical. Then, as shown
by Golse \cite{Golse2003} and Bardos et al. \cite{BardosGolseMauser2000,BardosErdosGolseMauserYau2002},
in the limit $N\rightarrow\infty$, the generation of correlations
between particles in time indeed gets suppressed (in the quantum BBGKY
hierarchy corresponding to (132-135)), and the time-dependent function
$\chi(\mathbf{q},t)$ satisfies (130-131). Likewise for the electrostatic
analogues of (132-133). Furthermore, we note that (132-135) is equivalent
to (109-116) by virtue of the Madelung transformation. \footnote{Presumably, then, there exists a Madelung BBGKY hierarchy corresponding
to (109-116), for which one can rigorously prove that in the limit
$N\rightarrow\infty$ the mean-field Madelung equations (119-123)
are recovered. We are unaware of such a proof in the mathematical
physics literature, however. }

Now, using the solution of (130-131), we can calculate the mean trajectory
of a typical \emph{zbw} particle at position $\mathbf{q}$ through
the mean equations of motion
\begin{equation}
\frac{d\mathbf{q}(t)}{dt}=\frac{\nabla S(\mathbf{q},t)}{m}|_{\mathbf{q}=\mathbf{q}(t)}=\frac{\hbar}{m}\mathrm{Im}\frac{\nabla\chi(\mathbf{q},t)}{\chi(\mathbf{q},t)}|_{\mathbf{q}=\mathbf{q}(t)},
\end{equation}
 
\begin{equation}
\begin{split}m\frac{d^{2}\mathbf{q}(t)}{dt^{2}}=\left[\partial_{t}\nabla S(\mathbf{q},t)+\frac{\nabla S(\mathbf{q},t)}{m}\cdot\nabla\left(\nabla S(\mathbf{q},t)\right)\right] & |_{\mathbf{q}=\mathbf{q}(t)}=-\nabla\left[m\Phi_{g}^{m.f.}(\mathbf{q},t)-\frac{\hbar^{2}}{2m}\frac{\nabla^{2}\sqrt{|\chi(\mathbf{q},t)|)}}{\sqrt{|\chi(\mathbf{q},t)|}}\right]|_{\mathbf{q}=\mathbf{q}(t)},\end{split}
\end{equation}
as well as the forward/backward stochastic trajectory through the
stochastic equations of motion

\begin{equation}
d\mathbf{q}(t)=\left[\frac{\hbar}{m}\mathrm{Im}\frac{\nabla\chi(\mathbf{q},t)}{\chi(\mathbf{q},t)}+\frac{\hbar}{m}\mathrm{Re}\frac{\nabla\chi(\mathbf{q},t)}{\chi(\mathbf{q},t)}\right]|_{\mathbf{q}=\mathbf{q}(t)}dt+d\mathbf{W}(t),
\end{equation}

\begin{equation}
d\mathbf{q}(t)=\left[\frac{\hbar}{m}\mathrm{Im}\frac{\nabla\chi(\mathbf{q},t)}{\chi(\mathbf{q},t)}-\frac{\hbar}{m}\mathrm{Re}\frac{\nabla\chi(\mathbf{q},t)}{\chi(\mathbf{q},t)}\right]|_{\mathbf{q}=\mathbf{q}(t)}dt+d\mathbf{W}_{*}(t).
\end{equation}
Considering that (130-131) is the leading-order large N approximation
to (132-135), trajectories calculated from (136-139) are expected
to only very roughly agree with the exact trajectories calculated
using the solutions of (132-135), whether for a dilute gas or plasma
of identical \emph{zbw} particles. Of course, in practice, it is impossible
to show this explicitly as it is a non-trivial problem to numerically
solve the system (132-133), even for just two particles.

Nonetheless, we can improve the mean-field approximation to (132-133)
by including the next-order terms in the large N limit. This was recently
done by us in \cite{DerakNewtLimStoGra2017} by (i) taking the Newtonian
limit of the Einstein-Langevin equation of semiclassical stochastic
gravity \cite{Hu2008}, and (ii) directly reconstructing the next-order
terms for the mean-field SN equations. The resulting `mean-field stochastic
SN equations' read 
\begin{equation}
i\hbar\frac{\partial\chi(\mathbf{q},t)}{\partial t}=\left[-\frac{\hbar^{2}}{2m}\nabla^{2}+m\Phi_{g}^{m.f.+}\right]\chi(\mathbf{q},t),
\end{equation}
\begin{equation}
\nabla^{2}\Phi_{g}^{m.f.+}=4\pi\left[m|\chi(\mathbf{q},t)|^{2}+\frac{\xi(\mathbf{q},t)}{2c^{2}}\right],
\end{equation}
\begin{equation}
\begin{split}<\xi(\mathbf{q},t)>_{s}=0, & \qquad<\xi(\mathbf{q}_{A},t_{A})\xi(\mathbf{q}_{B},t_{A})>_{s}=N(\mathbf{q}_{A},\mathbf{q}_{B};t_{A},t_{B}),\end{split}
\end{equation}
\begin{equation}
N(\mathbf{q}_{A},\mathbf{q}_{B};t_{A},t_{B})\coloneqq\mathrm{Re}\left\{ m^{2}c^{4}\chi^{*}(\mathbf{q}_{A},t_{A})\chi(\mathbf{q}_{B},t_{B})\delta^{3}\left(\mathbf{q}_{A}-\mathbf{q}_{B}\right)\delta\left(t_{A}-t_{B}\right)-m^{2}c^{4}|\chi(\mathbf{q}_{A},t_{A})|^{2}|\chi(\mathbf{q}_{B},t_{B})|^{2}\right\} .
\end{equation}

The bilocal field $N(\mathbf{q}_{A},\mathbf{q}_{B};t_{A},t_{B})$
is known as the ``noise kernel'', and essentially serves as a measure
of small (i.e., Gaussian) `quantum fluctuations' of the mass-energy
density of the N-particle system, as described by (142-143), between
two nearby space-time points $\left\{ \mathbf{q}_{A},t_{A}\right\} $
and $\left\{ \mathbf{q}_{B},t_{B}\right\} $. (Technically, the noise
kernel defined by (142-143) is divergent due to the spatial delta
function. This can be remedied by replacing the delta function with
a smearing function \cite{AnastopoulosHu2014,Anastopoulos2015}, but
for our purposes this detail is inessential.) Furthermore, the noise
kernel plays the role of the diffusion coefficient for the classical
stochastic (colored) noise field $\xi(\mathbf{q},t)$ (where $<...>_{s}$
refers to the statistical average), the latter of which phenomenologically
models the back-reaction of the quantum fluctuations on the gravitational
field via $\Phi_{g}^{m.f.+}$. \footnote{The fact that the noise field is colored instead of white implies
that $\xi(\mathbf{q},t)$ is a smooth function, which further implies
that the solution of (140-141) is a smooth function. } In other words, the noise field in (141) reincorporates the quantum
coherence of the gravitational potential to first-order in the large
N approximation. To see this last point more explicitly, we can observe
that the stochastic correction to $\Phi_{g}^{m.f.}$ 
\begin{equation}
\Phi_{g}^{+}(\mathbf{q},t)=-\frac{1}{c^{2}}\int d^{3}\mathbf{q}'\frac{\xi(\mathbf{q}',t)}{2|\mathbf{q}-\mathbf{q}'|},
\end{equation}
is known \cite{Hu2008} to formally reproduce the symmetrized two-point
correlation function for the quantized gravitational potential: \footnote{Equation (145) is deduced as follows. Start from the equality $2<h_{ab}(x_{A})h_{cd}(x_{B})>_{s}\begin{gathered}=\left\langle \Psi\right|\left\{ \hat{h}_{ab}(x_{A}),\hat{h}_{cd}(x_{B})\right\} \left|\Psi\right\rangle \end{gathered}
$ , where $h_{ab}(x_{A})$ is the classical stochastic metric perturbation
at spacetime point $x_{A}$ satisfying the regularized Einstein-Langevin
equation (see equation (3.14) of \cite{Hu2008}), $\hat{h}_{ab}(x_{A})$
is the quantum metric perturbation operator in the theory of perturbatively
quantized gravity (which is equivalent to the weak-field limit of
covariant path integral quantum gravity), and $\left|\Psi\right\rangle $
is the quantum state for a quantum field $\hat{\phi}(x)$ in the large
N expansion of covariant path integral quantum gravity \cite{HartleHorowitz1981,HuRouraVerdaguer2004,Hu2008}.
Implement the Newtonian limit by assuming $v\ll c$, $g_{ab}=\eta_{ab}+\delta\eta_{ab}$,
and $1\gg|T_{00}|/|T_{ij}|$; thus $\Phi_{g}^{+}\coloneqq\frac{1}{2}h_{00}$
and $\hat{\Phi}_{g}\coloneqq\frac{1}{2}\hat{h}_{00}$. Finally, project
$\left|\Psi\right\rangle $ onto the coherent state with corresponding
complex field $\chi$ \cite{DerakNewtLimStoGra2017}. (We show in
\cite{DerakNewtLimStoGra2017} that the coherent state projection
of $\left|\Psi\right\rangle $ is equivalent to the large N limit
of the many-body wavefunction $\psi$ of the exact Newtonian quantized-gravitational
level of description.) The result is (145).}
\begin{equation}
<\Phi_{g}^{+}(\mathbf{q}_{A},t_{A})\Phi_{g}^{+}(\mathbf{q}_{B},t_{B})>_{s}\begin{gathered}=\frac{1}{2}\left\langle \chi\right|\left\{ \hat{\Phi}_{g}(\mathbf{q}_{A},t_{A}),\hat{\Phi}_{g}(\mathbf{q}_{B},t_{B})\right\} \left|\chi\right\rangle \end{gathered}
.
\end{equation}

We say ``formally'' because the non-linear evolution (140-141) implies
failure of the Born-rule interpretation for $\chi$. Thus the `expectation
value' of the rhs of (145) cannot be understood as the standard quantum
expectation value. However, since $\chi$ does have a consistent stochastic
mechanical statistical interpretation (namely, $|\chi|^{2}$ corresponds
to the number density of \emph{zbw} particles at 3-space point $\mathbf{q}$
at time \emph{t}), we can ascribe a stochastic mechanical statistical
interpretation to the rhs of (145), in the sense that it is equivalent
(by the Madelung transformation) to the stochastic mechanical correlation
function: 
\begin{equation}
\left\langle \chi\right|\left\{ \hat{\Phi}_{g}(\mathbf{q}_{A},t_{A}),\hat{\Phi}_{g}(\mathbf{q}_{B},t_{B})\right\} \left|\chi\right\rangle =2\int_{-\infty}^{t}dt_{B}\int_{\mathbb{R}^{3}}d^{3}\mathbf{q}_{B}\rho(\mathbf{q}_{B},t_{B})\Phi_{g}(\mathbf{q}_{A},t_{A})\Phi_{g}(\mathbf{q}_{B},t_{B}),
\end{equation}
where $\Phi_{g}(\mathbf{q}_{A},t_{A})$ and $\Phi_{g}(\mathbf{q}_{B},t_{B})$
are solutions of the mean-trajectory Poisson equation (72).

Accordingly, if we use the solution of (140) in (136-139), the resulting
trajectories should slightly better approximate the exact trajectories
obtained from using the solutions of (132) for very large but finite
N. Note that with the solution of (140), the trajectories constructed
from integrating (136-137) contain classical (non-Markovian) stochastic
fluctuations through the stochasticity of the solution of (140). On
the other hand, the trajectories constructed from integrating (138-139)
contain classical stochastic fluctuations through the solution of
(140) \emph{and} the (Markovian) stochasticity encoded in the Wiener
process $d\mathbf{W}$ ($d\mathbf{W}_{*}$). Note, also, that even
though (140-143) are formulated for the case of a dilute system of
gravitationally interacting particles, they can also be applied to
dilute systems of electrostatically interacting particles, simply
by replacing $m\Phi_{g}^{m.f.+}\rightarrow e\Phi_{c}^{m.f.+}$ in
(140), which implies the replacements $\xi/c^{2}\rightarrow-\xi/c$
in (141) and $m^{2}c^{4}\rightarrow e^{2}c^{2}$ in (143). Then the
`stochastic mean-field Schr{\"o}dinger-Coulomb equations' provide a next-order
correction to the large N limit of the electrostatic analogue of (130-131),
and thereby partially reincorporate the quantum coherence of the N-particle
electrostatic potential operator. This last point can be seen most
explicitly by observing that (145) holds in the electrostatic case
as well, when we replace $\Phi_{g}^{+}\rightarrow\Phi_{c}^{+}$ and
$\hat{\Phi}_{g}\rightarrow\hat{\Phi}_{c}$. 

Finally, let us comment on the limitations of the mean-field approximations
considered here. 

First, the large N limit leading to (119-121) or (130-131) is only
applicable when the inter-particle interactions are sufficiently weak
that the independent-particle approximation is plausible. Some example
applications of (130-131) to self-gravitating N-particle systems that
conform reasonably well to the independent-particle approximation,
are boson stars \cite{Ruffini1969,Guzman2003,Liebling2012} and (when
one includes short-range interactions between particles) Bose-Einstein
condensates \cite{Chavanis2012,Chavanis2016}; for electrostatically
self-interacting N-particle systems, the electrostatic analogue of
(130-131) is widely used in condensed matter physics to model `jellium'
(i.e., homogeneous electron gas) systems \cite{Broglia2004,Giuliani2005}.
On the other hand, for strongly interacting N-particle systems such
as (say) superconducting microspheres \cite{RomeroIsart2012,Pino2016,DerAnaHu2016,DerakProbingGravCat2016},
the independent-particle approximation is a poor one and the deterministic
or stochastic SN/SC equations cannot be used.

Second, even for dilute N-particle systems, such as considered above,
the mean-field approximations provided by (130-131) and (140-143)
become empirically inadequate for calculating the gravitational force
on an external (macroscopic or mesoscopic or microscopic) probe, when
quantum fluctuations of the mass-energy density of the N-particle
system become too large. As an example, for the dilute system of N
gravitationally interacting ZSM particles, with total mass $M=Nm$,
suppose that the solution of (130) or (140) takes the form of a Schr{\"o}dinger
cat state. In particular, an equal-weighted superposition of two identical
Gaussians, where one is peaked at $\frac{1}{2}\mathbf{L}$, the other
at $-\frac{1}{2}\mathbf{L}$, and both having zero mean momentum:
\begin{equation}
\chi_{cat}(\mathbf{x})=\frac{1}{\sqrt{2}}\left[\chi_{left}(\mathbf{x})+\chi_{right}(\mathbf{x})\right]=\frac{1}{\sqrt{2}}\frac{1}{\left(2\pi\sigma^{2}\right)^{3/4}}\left[e^{-\frac{(\mathbf{x}+\mathbf{L}/2)^{2}}{4\sigma^{2}}}+e^{-\frac{(\mathbf{x}-\mathbf{L}/2)^{2}}{4\sigma^{2}}}\right].
\end{equation}
Then the Poisson equation for the mass density corresponding to (131)
or (141) takes the form

\begin{equation}
\nabla^{2}\Phi_{g}^{m.f.}=4\pi M|\chi(\mathbf{x})|^{2}=4\pi\left[\frac{M}{2}|\chi_{left}|^{2}+\frac{M}{2}|\chi_{right}|^{2}\right],
\end{equation}
or

\begin{equation}
\nabla^{2}\Phi_{g}^{m.f.+}=4\pi M|\chi(\mathbf{x})|^{2}=4\pi\left[\frac{M}{2}|\chi_{left}|^{2}+\frac{M}{2}|\chi_{right}|^{2}+\frac{\xi(\mathbf{x},0)}{2c^{2}}\right],
\end{equation}
with
\begin{equation}
\begin{split}<\xi(\mathbf{x},t)>_{s}=0, & \qquad<\xi(\mathbf{x}_{A},t_{A})\xi(\mathbf{x}_{B},t_{A})>_{s}=N(\mathbf{x}_{A},\mathbf{x}_{B};t_{A},t_{B}),\end{split}
\end{equation}
\begin{equation}
\begin{split}N(\mathbf{x}_{A},\mathbf{x}_{B};t_{A},t_{B}) & =\mathrm{Re}\left\{ M^{2}c^{4}\chi_{cat}^{*}(\mathbf{x}_{A},t_{A})\chi_{cat}(\mathbf{x}_{B},t_{B})\delta^{3}\left(\mathbf{x}_{A}-\mathbf{x}_{B}\right)\delta\left(t_{A}-t_{B}\right)-M^{2}c^{4}|\chi_{cat}(\mathbf{x}_{B},t_{B})|^{2}|\chi_{cat}(\mathbf{x}_{A},t_{A})|^{2}\right\} .\end{split}
\end{equation}
If the spatial separation between the two Gaussians is macroscopic,
e.g., $\mathbf{L}=1m$, and if $M=1,000kg$, then the classical gravitational
field produced by (148) or (149-151) is totally unrealistic. For example,
a probe corresponding to a macroscopic test mass passing through the
mid-point of the two mass distributions will, according to (148),
go undeflected, or, according to (149-151), will oscillate in between
the two mass distributions before passing through with no mean deflection
(because of the Gaussian property of the noise field). Both predictions
are in stark contrast to what the exact N-particle description (132-133)
would predict if $\psi(q)$ takes the form of (147) and one applies
the textbook quantum measurement postulates \cite{Bohm1951,BellAgainstMeasure}
or the stochastic mechanical theory of measurement \cite{Blanchard1986,Goldstein1987,Jibu1990,Blanchard1992,Peruzzi1996};
namely, that the test mass will either deflect towards the left mass
distribution or the right mass distribution, with probability $\frac{1}{2}$
each. \footnote{Of course, the stochastic SN equations (and the Einstein-Langevin
equation more generally) are formulated to handle only dilute N-particle
systems with small quantum fluctuations in the matter sector. Cat
state solutions clearly fall out of this regime, so it is not surprising
that the stochastic SN equations make an empirically inadequate prediction
in this case. In order to extend the stochastic SN equations to the
case of non-Gaussian fluctuations, we would (presumably) need to incorporate
into (140-141) the quantum coherence of the full n-point correlation
function involving $\hat{\Phi}_{g}$, in terms of some suitable generalization
of the noise kernel. This remains an open problem \cite{HuRouraVerdaguer2004,Hu2008,DerakNewtLimStoGra2017}.} Furthermore, apart from the fact that the solutions of (148) or (149-151)
don't have consistent Born-rule interpretations \cite{Adler2007,vanWezel2008,Hu2014,Derakhshani2014,Diosi2016},
the stochastic mechanical statistical interpretation of the solutions
of (148) or (149-151) doesn't predict a probed gravitational field
that's any more consistent with the prediction obtained from (132-133).
And, of course, all these issues with cat states apply as well in
the electrostatic case. 

As we will see in Part II, the limitations of the mean-field approximations
considered above can be circumvented by employing a center-of-mass
description of a large N system of ZSM-Newton/Coulomb particles. But
next let us compare ZSM-Newton/Coulomb, developed thus far, to other
semiclassical theories.

\section{Comparison to other semiclassical Newtonian field theories}

Here we compare ZSM-Newton/Coulomb, developed thus far, to other semiclassical
Newtonian field theories proposed in the literature. In particular,
we highlight conceptual advantages of the ZSM-Newton/Coulomb approach
and possibilities for experimental discrimination.

\subsection{Comparison to non-hidden-variable approaches}

Anastopoulos and Hu (AH) \cite{AnastopoulosHu2014} have shown that
the mean-field SN equations (130-131) can be derived from the standard
quantum field theoretic description of a scalar matter field interacting
with perturbatively quantized gravity (hereafter PQG): simply take
the Newtonian limit of PQG to obtain the N-particle Schr{\"o}dinger equation
(67), consider the case of weakly-coupled systems of identical particles,
then apply the large N limit (as we did in (132-135)). Complementing
their analysis, we have shown \cite{DerakNewtLimStoGra2017} that
the mean-field SN equations follow from standard semiclassical Einstein
gravity (SCEG) \cite{Hu2008,Hu2014,AnastopoulosHu2014}, under the
following prescription: (i) take the Newtonian limit of the semiclassical
Einstein equation (see (155) below) to obtain the Poisson equation
with the quantum expectation value of the mass density operator as
a source; (ii) introduce the Schr{\"o}dinger equation in Fock space for
the state-vector $\left|\psi\right\rangle $, with a gravitational
interaction term in the Hamiltonian involving the solution of the
Poisson equation; (iii) identify many-body quantum states consistent
with the large N regime of Newtonian PQG (i.e., coherent states);
(iv) compute the quantum expectation value of the mass density operator
with respect to such a state; and (v) project the Fock space Schr{\"o}dinger
equation into the first-quantized position space representation.

Likewise AH have shown \cite{AnastopoulosHu2014} that the mean-field
SC equations follow from standard relativistic QED: take the non-relativistic
limit, consider a weakly-coupled system of identical particles, then
take the large N limit. As with the gravitational case, we have also
shown \cite{DerakNewtLimStoGra2017} that the mean-field SC equations
follow from analogously applying steps (i-v) to standard semiclassical
relativistic electrodynamics (SCRED). \footnote{The semiclassical Maxwell equation of SCRED is given by $\nabla_{\mu}F^{\mu\nu}=\left\langle \psi\right|\hat{J}^{\nu}\left|\psi\right\rangle $,
where $\hat{J}^{\nu}$ is the charge four-current operator, $\left|\psi\right\rangle $
is some state-vector, and $\nabla_{\mu}$ is the covariant derivative
in case the background spacetime is curved. Taking the non-relativistic
limit, introducing the Fock space Schr{\"o}dinger equation for $\left|\psi\right\rangle $,
and taking $\psi$ to be a coherent state, one obtains the mean-field
SC system \cite{DerakNewtLimStoGra2017}. }. 

Thus, for weakly-coupled systems of identical particles, the large
N limit scheme used in ZSM-Newton/Coulomb can also be employed in
Newtonian PQG/QED; and in both cases one recovers the mean-field SN/SC
equations. These results also agree with the Newtonian limits of SCEG
and SCRED, when the latter are interpreted as mean-field theories
for weakly-coupled systems of identical particles.

It is remarkable that these correspondences follow despite ZSM-Newton/Coulomb
treating the gravitational/Coulomb potentials as fundamentally classical
fields sourced by point-like classical particles undergoing non-classical
motions in 3-space. In this respect, the ZSM approach is unique among
existing formulations of quantum theory that have been extended to
fundamentally-semiclassical gravity or electrodynamics.

For example, it is well known \cite{Diosi1984,Salzman2005,Carlip2008,Derakhshani2014,AnastopoulosHu2014,BahramiBassi2014a,Giulini2014,Bahrami2015,DerakNewtLimStoGra2017}
that if one formulates fundamentally-semiclassical gravity based on
the equations of either standard non-relativistic quantum mechanics
\cite{Diosi1984,Salzman2005,Carlip2008,Derakhshani2014,AnastopoulosHu2014,BahramiBassi2014a,Giulini2014,Bahrami2015,DerakNewtLimStoGra2017}
or non-relativistic many-worlds interpretations \cite{Yang2013,Derakhshani2014},
one obtains the \emph{N}-body SN equations

\begin{equation}
i\hbar\frac{\partial\psi(q,t)}{\partial t}=\sum_{i=1}^{N}\left[-\frac{\hbar^{2}}{2m_{i}}+\frac{m_{i}\Phi_{g}^{SN}}{2}\right]\psi(q,t),
\end{equation}
and 
\begin{equation}
\nabla^{2}\Phi_{g}^{SN}=4\pi m(\mathbf{q},t)=4\pi\underset{i=1}{\overset{N}{\sum}}\int d^{3}\mathbf{r}{}_{1}...d^{3}\mathbf{r}{}_{N}|\psi(\mathbf{r}_{1}...\mathbf{r}_{N},t)|^{2}m_{i}\delta^{(3)}(\mathbf{q}-\mathbf{r}{}_{i}),
\end{equation}
where 
\begin{equation}
\Phi_{g}^{SN}=-\sum_{j=1}^{N(j\neq i)}\int\frac{m_{j}(\mathbf{q}'_{j},t)}{|\mathbf{q}_{i}-\mathbf{q}'_{j}|}d^{3}\mathbf{q}'{}_{1}...d^{3}\mathbf{q}'_{N}.
\end{equation}
It is also well-known \cite{Derakhshani2014,AnastopoulosHu2014,BahramiBassi2014a,Giulini2014,DerakNewtLimStoGra2017}
that (152-153) can be obtained from the Newtonian limit of the semiclassical
Einstein equation
\begin{equation}
G_{nm}=\kappa\left\langle \psi\right|\hat{T}_{nm}\left|\psi\right\rangle ,
\end{equation}
\emph{if} one naively assumes that (155) is valid even when $\psi$
is a single-particle wavefunction, whether in a standard quantum theory
reading or a many-worlds interpretation (re: the latter context, see
\cite{PageGeilker1981,Yang2013,Derakhshani2014}). However, like the
mean-field SN equations, the solutions of (152-153) lack consistent
Born-rule interpretations \cite{Adler2007,vanWezel2008,Hu2014,Derakhshani2014,Diosi2016}
and include the macroscopic gravitational cat states discussed in
section 4. In other words, attempting to formulate fundamentally-semiclassical
gravity, based on either standard quantum theory or many-worlds interpretations,
results in a nonlinear classical-gravitational field theory that makes
absurd empirical predictions.As another example, it was shown in \cite{Derakhshani2014,DerakProbingGravCat2016}
that the \emph{N}-body SN equations (with stochastic corrections to
dynamically induce intermittent wavefunction collapse) arise naturally
when one extends the GRW, CSL, and DP theories to fundamentally-semiclassical
gravity with a matter density ontology (called GRWmN, CSLmN, and DPmN,
respectively). In contrast to SQM-Newton (where SQM = standard quantum
mechanics) and MW-Newton (where MW = many worlds), GRWmN/CSLmN/DPmN
have been shown to adequately suppress the empirically problematic
macroscopic gravitational cat states while also having consistent
statistical interpretations \cite{Derakhshani2014,DerakProbingGravCat2016}.
Thus, these dynamical collapse theories of fundamentally-semiclassical
Newtonian gravity are empirically viable. At the same time, these
dynamical collapse theories also make slightly different empirical
predictions from the Newtonian large N limit of PQG and SCEG; and
given the empirical equivalence between Newtonian-large-N PQG and
SCEG, and N-particle ZSM-Newton (when the nonlinear terms of the latter
are neglected), it will also be the case that these dynamical collapse
theories make slightly different empirical predictions from N-particle
ZSM-Newton. These slight differences in empirical predictions are
entailed by the collapse-inducing stochastic correction terms, and
the fact that these dynamical collapse theories still allow for stable
gravitational cat states in a mesoscopic regime of masses \cite{Derakhshani2014,DerakProbingGravCat2016}.
The slightly different empirical predictions of these collapse theories
may be testable by the next (or next-next) generation of state-of-the-art
AMO experiments, as argued by us in \cite{DerAnaHu2016,DerakProbingGravCat2016}. 

As yet another example, the Tilloy-Di{\'o}si (TD) model of fundamentally-semiclassical
gravity makes use of the flash ontology within CSL or DP dynamics,
to describe fundamentally-semiclassical Newtonian gravitational interactions
between \emph{N} particles, with no nonlinear feedback from the wavefunction.
(One can also make a GRW analogue of the TD model, as pointed out
by us in \cite{Derakhshani2014}.) TD's (stochastic) analogue of the
SN equations reads

\begin{equation}
\begin{aligned}\frac{d\left|\psi\right\rangle }{dt} & =-\frac{i}{\hbar}\left(\hat{H}+\hat{V}_{G}\right)\left|\psi\right\rangle \\
 & -\frac{1}{8\pi\hbar G}\int d\mathbf{r}\left(\nabla\hat{\Phi}(\mathbf{r})-\left\langle \nabla\hat{\Phi}(\mathbf{r})\right\rangle \right)^{2}\left|\psi\right\rangle \\
 & -\hbar\left(1+i\right)\int d\mathbf{r}\left(\hat{\Phi}(\mathbf{r})-\left\langle \nabla\hat{\Phi}(\mathbf{r})\right\rangle \right)\delta\rho(\mathbf{r})\left|\psi\right\rangle ,
\end{aligned}
\end{equation}
up to a fixed spatial cut-off $\sigma$. Here the potential $\hat{V}_{G}$
represents the usual Newtonian gravitational potential operator, while
the non-Hermitian terms on the right give rise to decoherence and
collapse of spatial superpositions of a massive particle. As shown
by TD \cite{Tilloy2016}, their model adequately suppresses macroscopic
gravitational cat states and has a consistent statistical interpretation.
By virtue of the non-Hermitian terms in (156), the TD model also makes
slightly different predictions from both Newtonian-limited PQG and
ZSM-Newton. These differences might also be testable by the next (or
next-next) generation of state-of-the-art AMO experiments \cite{DerakProbingGravCat2016}.
A notable conceptual difference between the TD model and ZSM-Newton
is that the former predicts point-like mass distributions (which source
the classical gravitational field) that discontinuously appear and
disappear in space-time, because the flash ontology is used as the
means of defining the mass density sources (we have previously made
this point in regards to a GRW analogue of the TD model \cite{Derakhshani2014});
by contrast, the mass density sources in ZSM-Newton (the \emph{zbw}
particles) involve no such discontinuities. 

Concerning theories of fundamentally-semiclassical electrodynamics,
perhaps the best-known is Asim Barut's ``self-field QED'' \cite{Barut1988a,Barut1988b,BarutDowling1990a,BarutDowling1990b}.
This theory essentially takes the Schr{\"o}dinger-Coulomb (SC) analogue
of (152-153) (or its relativistic generalization, the Dirac-Maxwell
system) as its starting point and purports to reproduce the self-energy
effects of non-relativistic and relativistic QED to all orders of
perturbation linear in alpha. However, there are more basic predictions
of the theory that were left (apparently) unaddressed by Barut and
his co-workers, and which seem to make the theory empirically inadequate.
First, just like the SN equations, the SC analogue of (152-153) does
not have a consistent Born-rule interpretation, thereby preventing
a naive application of the standard quantum measurement postulates.
Second, also just like the SN equations, the SC equations admit macroscopic
electrostatic cat states as solutions, and these solutions are clearly
not seen in the real world (incidentally, this rules out the possibility
of many-worlds interpretations based on the SC equations). Third,
even if one attempts to add stochastic corrections to the SC equations
in the form of GRW/CSL/DP, numerical simulations of the SC equations
indicate that a free particle wavepacket would undergo Coulomb self-repulsion
(from the nonlinear electrostatic self-interaction), and this self-repulsion
effect would lead to interference maxima in the two-slit experiment
much too broad to be in agreement with existing experimental data
\cite{GrossardtPersonalComm2013}. As an alternative formulation of
fundamentally-semiclassical electrodynamics based on dynamical collapse
theories, we might consider a straightforward electrostatic analogue
of TD's equation (156). Presumably such a theory would be free of
the problems entailed by the nonlinearity of the SC equations, but
this remains to be explored. In any case, it would appear that, in
comparison to theories of fundamentally-semiclassical electrodynamics
based on standard quantum mechanics, many-worlds interpretations,
and dynamical collapse theories (with matter density ontology), ZSM-Coulomb
is the only one that's empirically viable (within its non-relativistic
domain of validity) insofar as it's empirically equivalent to the
Newtonian limits of standard QED and SCRED (modulo the tiny empirical
differences entailed by the nonlinear correction terms (86-90) discussed
in section 3).

\subsection{Comparison to alternative hidden-variable approaches}

Other formulations of stochastic mechanics exist besides ZSM \cite{Fenyes1952,Davidson1979,Yasue1981a,Wallstrom1989,Wallstrom1994,Derakhshani2016a,Derakhshani2016b}.
Moreover, dBB pilot-wave theory is the most well-developed hidden-variables
formulation of quantum theory to date. Do these other hidden-variables
theories have consistent and empirically adequate extensions to semiclassical
Newtonian field theories, whether in the form of fundamentally-semiclassical
theories or semiclassical approximations? How do they compare and
contrast to ZSM-Newton/Coulomb?

As mentioned in section 2, all non-ZSM formulations of stochastic
mechanics are subject to Wallstrom's criticism \cite{Wallstrom1989,Wallstrom1994,Bacciagaluppi2005,Bacciagaluppi2012,Derakhshani2016a,Derakhshani2016b}
- they are all empirically inadequate because they either allow for
too many solutions or too few solutions, compared to the Schr{\"o}dinger
equation of standard quantum mechanics. For those formulations that
allow too many solutions, one can always impose by hand the quantization
condition needed in order to make the solution spaces of those formulations
isomorphic to the solution space of standard quantum mechanics \cite{Wallstrom1989,Wallstrom1994,Bacciagaluppi2005,Bacciagaluppi2012,Derakhshani2016a,Derakhshani2016b}.
This is, of course, an ad hoc move, but one might view it as provisional
until such a condition can be justified by some non-ZSM modification
of said formulations of stochastic mechanics. In this case, the amended
formulations of stochastic mechanics would result in exactly the same
mathematical descriptions of Newtonian gravity and electrodynamics
as we've found for ZSM, both at the exact (i.e., N-particle Schr{\"o}dinger
equation) level and the level of the mean-field approximation schemes.
(Differences would arise, however, in physically motivating the mean-field
approximation, e.g., ansatz (111) in section 4; since the $S$ function
would not be interpretable as the phase of a periodic phenomenon localized
to the stochastic mechanical particle, such an ansatz would have to
be imposed ad hoc.)

Concerning semiclassical de Broglie-Bohm theories, let us consider
the possibilities separately.

\subsubsection{Comparison to fundamentally-semiclassical de Broglie-Bohm theories}

There is some ambiguity in how to construct a dBB-based theory of
fundamentally-semiclassical Newtonian gravity (or electrodynamics).
First, one has to make a choice about which version of dBB dynamics
to consider (i.e., the `first-order' or `second-order' version \cite{Bohm1952I,Bohm1952II,BohmHiley1993,HollandBook1993,Duerr2009,DGZbook2012,OriolsMompart2012,Goldstein2013}).
Second, given a version, one has to make a choice about how to interpret
its ontology (e.g., is the wavefunction part of the ontology or does
it merely play a `nomological' role in the theory?). Third, one has
to make a choice about which part of the dBB ontology - the wavefunction
or the particles - plays the role of the mass (or charge) density
that sources the classical gravitational (or electromagnetic) field;
as it turns out, for versions of dBB in which the particles don't
constitute the only ontic variables, there is no compelling reason
why the particles (as opposed to the wavefunction) should be the mass
(charge) density source for the classical gravitational (electromagnetic)
field, even though that might seem like a prima facie natural choice. 

Let us consider this last point in more detail for the gravitational
case first, under the first-order `dual space' version of non-relativistic
dBB \cite{Bohm1952I,Bohm1952II,BellQMCosmo2004,HollandBook1993,BohmHiley1993,Goldstein2013}.
In other words, the version of dBB theory that posits an ontic 3N-dimensional
configuration space, occupied by an ontic `universal wavefunction'
$\psi(\mathbf{q}_{1},,,\mathbf{q}_{N},t)$, and an ontic 3-dimensional
space (which exists completely independently of the configuration
space) occupied by N (spinless) particles with configuration $q(t)=\left\{ \mathbf{q}_{1}(t),...,\mathbf{q}_{N}(t)\right\} $.
The universal wavefunction \footnote{The universal wavefunction is required to satisfy the usual boundary
conditions of single-valuedness, smoothness, and finiteness.} evolves by the Schr{\"o}dinger equation 

\begin{equation}
i\hbar\frac{\partial\psi}{\partial t}=\left[-\sum_{i=1}^{N}\frac{\hbar^{2}}{2m_{i}}\nabla_{i}^{2}+V^{int}\right]\psi,
\end{equation}
where $V^{int}$ is some scalar interaction potential to be specified
and we assume the normalization $\int_{\mathbb{R}^{3N}}|\psi|^{2}d^{3N}q=1$.
The particles evolve by the guiding equation
\begin{equation}
\frac{d\mathbf{q}_{i}(t)}{dt}=\frac{\hbar}{m_{i}}\mathrm{Im}\frac{\nabla_{i}\psi}{\psi}|_{\mathbf{q}_{j}=\mathbf{q}_{j}(t)}=\frac{\nabla_{i}S}{m_{i}}|_{\mathbf{q}_{j}=\mathbf{q}_{j}(t)},
\end{equation}
for all $i=1,..,N$, where the $\nabla S$ form follows if we write
$\psi=|\psi|e^{iS/\hbar}$. In addition, we have ``equivariance''
\cite{Duerr2009,OriolsMompart2012,Goldstein2013}, i.e., the statement
that if the initial particle configuration of the dBB system is distributed
as $\rho_{0}=|\psi_{0}|^{2}$, then this ``quantum equilibrium distribution''
\cite{Duerr2009,OriolsMompart2012,Goldstein2013} is preserved under
time-evolution by the quantum continuity equation implicit in (157).
In other words, the quantum continuity equation implicit in (157)
entails the map $|\psi_{0}|^{2}\rightarrow|\psi_{t}|^{2}$.

Notice that both the wavefunction and the particles `feel' the mass
parameters $\left\{ m_{1},...,m_{N}\right\} $. More specifically,
the time-evolution of $\psi$ (at every point in configuration space)
through (157) explicitly depends on all the mass parameters via the
kinetic energy operators, while the evolution of $\mathbf{q}_{i}(t)$
depends explicitly on only $m_{i}$ but implicitly on all the other
mass parameters through the positions of all the other particles.
The dependence of the evolution of $\psi$ on the mass parameters
is made even more manifest by starting from the N-particle Bohm-Dirac
theory \cite{Holland1992,HollandBook1993}, i.e., the most straightforward
relativistic N-particle extension of (157-158), and then taking the
non-relativistic limit; we would find that the positive-energy components
of the Dirac spinor in the Bohm-Dirac theory evolve by a corrected
version of (157), where the correction terms are rest-energy terms
$\sum_{i=1}^{N}m_{i}c^{2}$ in the Hamiltonian operator. 

One might think that since a classical gravitational field lives (by
definition!) in 3-space, and since only the particles live in 3-space,
this is why the particles should be the (point) sources for the gravitational
field. However, recall that the rhs of (153) gives a natural definition
of a 3-space mass density in terms of $\psi$ in configuration space. 

Consequently, it would seem that inertial mass is a property of both
the wavefunction and the particles, and there seems to be no justification
for assuming that the particles \emph{must} be used \emph{solely}
as the mass density sources for a classical gravitational field, if
one wants to make a fundamentally-semiclassical Newtonian gravitational
theory out of the present version of dBB. Not only that, if one allows
$\psi$ to have properties such as energy density, momentum density,
etc., one can define the Hamiltonian density 
\begin{equation}
\mathcal{H}=\psi^{\ast}\left[-\sum_{i=1}^{N}\frac{\hbar^{2}}{2m_{i}}\nabla_{i}^{2}+V^{int}+m_{i}c^{2}\right]\psi.
\end{equation}
This Hamiltonian density has the physical interpretation of the energy
density stored in the ontic wavefunction, and indicates that the rest-energy
terms, hence the $m_{i}$, compose the total mass-energy density of
the wavefunction in configuration space. To be sure, nothing in the
first-order version of dBB or the dual space version thereof \emph{requires}
that $\psi$ have additional properties like energy density; but nothing
excludes these additional properties either. In any case, \emph{if}
one allows $\psi$ to have properties like energy density, then the
present version of dBB theory seems to make a compelling case for
(at least) taking $\psi$ to be the mass density source for the classical
gravitational field. 

Given that the dBB theory under consideration is ambiguous about which
part of its ontology should be used (or is most natural to use) as
the mass density source for a classical gravitational field, let us
consider the empirical consequences of using either the wavefunction
or the particles or both.

If $\psi$ is used as a source, then the Poisson equation for the
classical gravitational field takes the SN form (153), and the Schr{\"o}dinger
equation (157) takes the SN form (152). Because of the conceptual
and technical problems with the SN system (152-153), i.e., inconsistency
with the Born-rule interpretation and prediction of macroscopic semiclassical
gravitational cat states, we must conclude that this version of fundamentally-semiclassical
dBB Newtonian gravity (hereafter, dBBfsc-Newton1 where ``fsc'' =
``fundamentally-semiclassical'') is not empirically viable.

If the particles are used as point sources, then the Poisson equation
takes the form 
\begin{equation}
\nabla^{2}\Phi_{g}=4\pi\sum_{i=1}^{N}m_{i}\delta^{3}\left(\mathbf{q}-\mathbf{q}_{i}(t)\right),
\end{equation}
where the $\mathbf{q}_{i}(t)$ are solutions of the guiding equation
(158) for all $i=1,..,N$. The solution of (160) then yields the inter-particle
gravitational potential energy, which depends on the actual positions
of all the dBB particles at a single time, and feeds back into the
Schr{\"o}dinger equation (157), giving

\begin{equation}
i\hbar\frac{\partial\psi}{\partial t}=\left[-\sum_{i=1}^{N}\frac{\hbar^{2}}{2m_{i}}\nabla_{i}^{2}+V_{g}^{int}(\mathbf{q}_{i}(t),\mathbf{q}_{j}(t))\right]\psi,
\end{equation}
where 
\begin{equation}
V_{g}^{int}(\mathbf{q}_{i}(t),\mathbf{q}_{j}(t))=\sum_{i=1}^{N}\frac{m_{i}\Phi_{g}}{2}=-\sum_{i=1}^{N}\frac{m_{i}}{2}\sum_{j=1}^{N(j\neq i)}\frac{m_{j}}{|\mathbf{q}_{i}(t)-\mathbf{q}_{j}(t)|}.
\end{equation}
This version of dBBfsc-Newton (dBBfsc-Newton2) was also considered
by Struyve \cite{Struyve2015}, who suggested that it might constitute
a viable alternative to the SN equations. (Kiessling considered the
electrostatic analogue in \cite{Kiessling2006}.) However, dBBfsc-Newton2
appears to be empirically inadequate since $|\psi_{t}|^{2}$ depends
now on the actual positions of all the particles at each time, implying
that the equivariance property breaks down (i.e., $|\psi(q(t),t)|^{2}$
does not satisfy the quantum continuity equation and can no longer
be interpreted as a probability density). The break-down of equivariance
means that dBBfsc-Newton2 makes no statistical predictions, whether
for gravitational or non-gravitational interactions at the Newtonian
level. In other words, even for, say, a non-gravitational position
measurement of the dBB particle, the density $|\psi_{t}|^{2}$ corresponding
to (161-162) doesn't have a consistent probabilistic interpretation,
whether as a distribution over an ensemble of identical fictitious
dBB particles \cite{HollandBook1993,OriolsMompart2012} or as a typicality
measure \cite{Duerr1992,Duerr2009,DGZbook2012,Goldstein2013}.

Using both the wavefunction and the particles as mass density sources
for classical gravitational fields, doesn't yield an empirical inadequate
theory either; such a theory (dBBfsc-Newton3) lacks the equivariance
property (hence making no statistical predictions), and the double
counting of the Newtonian gravitational field would lead to gross
inconsistency with the gravitational field predicted by classical
Newtonian gravity for macroscopic mass distributions.

We must therefore conclude that there does not appear to be an empirically
viable formulation of dBBfsc-Newton that's based on the first-order
dual-space version of dBB. Moreover, we do not see how to obtain an
empirically viable formulation of dBBfsc-Newton using other versions
of first-order dBB theory, whether Albert's `world particle' formulation
\footnote{In fact, it does not even seem possible to define a fundamentally-semiclassical
gravity theory using Albert's formulation. Albert's formulation takes
as fundamental ontological postulates (i) configuration space $\mathbb{R}^{3N}$,
(ii) $\psi$ in configuration space evolving by the N-particle Schr{\"o}dinger
equation, the latter defined in terms of a Hamiltonian that includes
an N-particle interaction potential $\hat{V}^{int}(\hat{\mathbf{q}}_{i},\mathbf{\hat{q}}_{j})$
that's written in a preferred coordinate system, and (iii) a single
configuration point (the world particle) in $\mathbb{R}^{3N}$, evolving
by the guidance equation. 3-space, and a configuration of particles
in 3-space, are claimed to be emergent ontologies in the sense that
they are claimed to arise from a philosophical-functionalist analysis
of $\hat{V}^{int}$ and the latter's influence on the motion of the
world particle through $\psi$. But a classical gravitational field
in 3-space is not part of the emergent ontology. And, of course, presupposing
a classical gravitational field living in 3-space is not allowed,
as that would contradict the entire purpose of Albert's formulation
(which is to regard configuration space, and the ontological variables
living in it, as the fundamental ontologies).} \cite{Albert2015}, Norsen's TELB formulation \footnote{The TELB (Theory with Exclusively Local Beables) formulation differs
from the dual space formulation in that 3-space is the only ontic
space. This approach is (mathematically) motivated by Taylor-expanding
$\psi$ in configuration space into an infinite hierarchy of nonlocally
coupled fields in 3-space; more precisely, each particle has a single-particle
wavefunction pushing it around via the guidance equation, but the
single-particle wavefunction is coupled to an infinite hierarchy of
3-space ``entanglement fields'', which are themselves nonlocally
coupled to the entanglement fields of every other particle (hence
why they are called ``entanglement'' fields). The postulate $\rho_{0}=|\psi_{0}|^{2}$
is still imposed on the single-particle wavefunctions, and equivariance
still holds. One could then define classical gravitational fields
directly in terms of mass density sources built out of the single-particle
wavefunctions, but this would just lead to a TELB version of the SN
equation, which would entail all the empirically problematic predictions
of the SN equation (e.g., macroscopic gravitational cat states). Furthermore,
the break-down of equivariance from using the particles as point sources
would still remain. } \cite{Norsen2010,Norsen2014}, or D{\"u}rr-Goldstein-Zangh{\`i}'s nomological
formulation \footnote{The nomological formulation is still conjectural, but the basic idea
is that the `fundamental' wavefunction is the time-independent Wheeler-DeWitt
wavefunctional $\Psi\left(h,\phi\right)$, interpreted as part of
\emph{physical law} rather than physical ontology. Time-dependent
wavefunctions are suggested to be derived, effective descriptions
for `subsystems' of the universe, and not part of physical ontology
either. Only 3-space and particles living in 3-space constitute physical
ontology. Accordingly, one cannot not use time-dependent wavefunctions
in the definition of an SN-type classical mass-density source in 3-space,
as this would be inconsistent with the expected Newtonian limit of
the Wheeler-DeWitt equation (i.e., the usual linear Schr{\"o}dinger equation
involving an operator-valued gravitational interaction potential)
\cite{DGZ1995,GoldsteinZanghi2011}. Nor could one use the dBB particles
as point sources for a classical gravitational field coupling back
to the time-dependent wavefunction, as this is inconsistent with the
expected Newtonian limits of the Wheeler-DeWitt equation and the guiding
equations for $h$ and $\phi$ (i.e., the equations of dBB-Newton).
So the nomological formulation of dBB also doesn't allow for a formulation
of fundamentally-semiclassical Newtonian gravity.} \cite{DGZ1995,Duerr2009,GoldsteinZanghi2011,DGZbook2012,Goldstein2013}.
Second-order formulations of dBB, namely the ``ontological interpretation''
advocated by Bohm-Hiley \cite{BohmHiley1993} and Holland \cite{HollandBook1993},
don't seem to change the situation either: their only difference from
first-order formulations of dBB is that the Schr{\"o}dinger equation
and wavefunction are replaced by the Madelung equations for $|\psi|$
and $S$, with the quantization condition imposed on the latter. 

By comparison, while the ontology of ZSM-Newton involves more than
just particles, it is clear from the very formulation of ZSM-Newton
that the particles must be understood as possessors of inertial mass.
This is manifest from (i) the definition of the rest-mass of a \emph{zbw}
particle as corresponding to the energy associated with the Compton
frequency oscillation of the \emph{zbw} particle in its rest frame,
and (ii) the definition of the \emph{i}-th Wiener process, which describes
the stochastic evolution of the \emph{i}-th particle position and
depends on the \emph{i}-th mass parameter through the diffusion coefficient
$\hbar/m_{i}$. Furthermore, as we argued in section 3, while the
ether of ZSM is expected to carry stress-energy, it is expected to
be negligible in the Newtonian regime. Thus, in contrast to dBB, ZSM
seems to make the choice of the particles as mass density sources
for a classical gravitational field, inevitable. Another difference
from dBB is the following: recall from section 3 that, because the
Schr{\"o}dinger equation and wavefunction are derived in ZSM, the use
of the particles as sources for a classical gravitational field doesn't
entail the nonlinear coupling in (161-162); rather, as we saw in section
3, the gravitational field that does couple to the Schr{\"o}dinger equation/wavefunction
corresponds (to leading order) to $\hat{V}_{g}^{int}(\mathbf{\hat{q}}_{i},\mathbf{\hat{q}}_{j})$.
This is why ZSM-Newton avoids a break-down of the equivariance property.
So despite ZSM and dBB sharing many equations in common - the Schr{\"o}dinger
equation (157), the guiding equation (158), and equivariance of $\rho_{0}=|\psi_{0}|^{2}$
- and despite both theories sharing in common a ``primitive ontology''
\footnote{Primitive ontology is defined by Allori et al. \cite{Allori2012}
as ``variables describing the distribution of matter in 4-dimensional
space-time''.} involving particles with definite 3-space trajectories, the different
axioms on which ZSM and dBB are based lead to significantly different
conclusions about how to formulate a theory of fundamentally-semiclassical
Newtonian gravity, and the empirical viability thereof. 

It is a straightforward exercise to demonstrate that analogous conclusions
follow from consideration of the electrodynamical case, i.e., dBBfsc-Coulomb
theories vs. ZSM-Coulomb. This being said, we wish to evaluate a well-known
peculiarity of standard dBB theory involving charge-field coupling
(i.e., dBB-Coulomb), from the viewpoint of ZSM-Coulomb. 

For a single-particle dBB system, in the presence of an external magnetic
vector potential $\mathbf{A}_{ext}(\mathbf{q},t)$, the momentum operator
in the Schr{\"o}dinger equation gets a correction $\mathbf{\hat{p}}\rightarrow\mathbf{\hat{p}}-e\mathbf{A}_{ext}$.
Now, consider the magnetic Aharonov-Bohm (AB) effect in dBB \cite{PhilippidisBohmKaye1982,BohmHiley1993,HollandBook1993},
where $\mathbf{A}_{sol}=\left(\Phi/2\pi r\right)\hat{\theta}$ is
the magnetic vector potential sourced by an infinitely long cylindrical
solenoid with flux $\Phi$. For an electron wavepacket split into
two partial packets passing on either side of the solenoid, where
the paths $P_{1}$ and $P_{2}$ traversed by the packets form a loop
$C$ encircling the solenoid, the correction to the momentum operator
entails a phase shift $\psi\rightarrow\psi'=N'\left[\psi_{1}+\psi_{2}e^{ie\Phi/\hbar}\right]e^{\left(ie/\hbar\right)\int_{P_{1}}\mathbf{A}_{sol}\cdot d\mathbf{q}}$,
when the packets are recombined to form an interference pattern ($N'$
is a normalization constant). Correspondingly, the position probability
density associated to the interference pattern gets shifted as $|\psi'|^{2}=\rho'=N'^{2}\left\{ \rho_{1}+\rho_{2}+2\sqrt{\rho_{1}}\sqrt{\rho_{2}}cos\left[(S_{1}-S_{2})/\hbar-\delta\right]\right\} $,
where $\delta=e\Phi/\hbar$. Note that while the dBB particle moves
along with only one of the packets around the solenoid, say the packet
traversing path $P_{1}$, with modified momentum $\mathbf{p}=\nabla S_{1}-e\mathbf{A}_{sol}$,
both packets `feel' $\mathbf{A}_{sol}$ since each picks up a phase
factor $\psi_{a}\rightarrow\psi_{a}e^{(ie/\hbar)\int_{P_{a}}\mathbf{A}_{sol}\cdot d\mathbf{q}}$
such that $\Phi=\oint_{C}\mathbf{A}_{sol}\cdot d\mathbf{q}=\int_{P_{1}}\mathbf{A}_{sol}\cdot d\mathbf{q}-\int_{P_{2}}\mathbf{A}_{sol}\cdot d\mathbf{q}$
and $\oint_{C}\mathbf{p}\cdot d\mathbf{q}=nh-e\Phi$. In other words,
even though the motion of the dBB particle is altered by the presence
of the vector potential, suggesting (seemingly) that the charge $e$
is a property localized to the dBB particle (like in classical electrodynamics),
the fact that the `empty' packet (i.e., the packet moving along $P_{2}$)
also picks up a phase factor, and that this phase factor contributes
to the shift in the interference pattern of the recombined packets,
suggests that charge is \emph{also} a property carried by the (spatially
delocalized) wavefunction \cite{PhilippidisBohmKaye1982,BohmHiley1993,HollandBook1993,BrownDewdneyHorton1995}.
A completely analogous situation arises for the gravitational analogue
of the magnetic AB effect, where $\mathbf{A}_{sol}$ is the gravitomagnetic
vector potential sourced by a solenoid carrying a mass (instead of
charge) current, and all other expressions are identical except for
the replacement $e\rightarrow m$ \cite{Hohensee2012}. Analogous
considerations apply to the case of the electric/gravitoelectric AB
effect.

Since the dBB treatment of the AB effect is formally the same as the
ZSM-Coulomb/Newton treatment of the AB effect, this might seem to
conflict with the ZSM-Coulomb/Newton hypothesis that the charge (rest
mass) of a system is a property localized to \emph{zbw} particles.
However, there is no inconsistency. In ZSM-Coulomb/Newton, the finding
that the empty packet in the AB effect picks up a phase factor that
contributes to the shift in the interference pattern intensity is
a \emph{consequence} of the following set of postulates: (i) rest-mass
and charge are intrinsic properties of \emph{zbw} particles; (ii)
the \emph{zbw} particles, whose oscillations are dynamically driven
by the ether medium, always have well-defined mean phases along their
3-space trajectories; and (iii) the diffusion process for the \emph{zbw}
particles in the ether satisfies the global constraint of being conservative.
It might then be asked if ZSM-Coulomb/Newton gives physical insight
into what it means, in terms of its proposed underlying ontological
picture of the world, for empty packets to electromagnetically (or
gravitationally) couple to external fields, even though it is the
\emph{zbw} particles that carry the rest-mass and charge of a system.
We can sketch an answer as follows.

As discussed in \cite{Schroedinger1938,Bohm1951,Derakhshani2016a},
the superposition principle for wavefunctions is a consequence of
the single-valuedness condition, and the single-valuedness condition
on wavefunctions in ZSM follows from the union of postulates (ii)
and (iii). And as we've discussed in \cite{Derakhshani2016b}, an
empty packet describes possible alternative histories of a Nelsonian/\emph{zbw}
particle through a different region of the ether (the different region
corresponding to the spatial support of the empty wavepacket in 3-space),
while also indirectly reflecting spatio-temporal variations in that
different region of the ether (because the ether-sourced osmotic potential
$U(q,t)$ changes as a function of space and time via the continuity
equation and is constrained by boundary conditions in the environment).
Thus the empty packet traversing path $P_{2}$ reflects (indirectly)
a region of the ether that's (spatio-temporally) varying along $P_{2}$,
and the interference of the recombined packets reflects (indirectly)
two regions of ether recombining and interfering while satisfying
postulates (ii) and (iii). Since the ether medium is presumed to pervade
all of 3-space, and since all components of the ether are presumed
to be nonlocally connected to each other, the ether region corresponding
to the empty packet is actually not physically independent of the
ether region corresponding to the occupied packet. In other words,
for the ether to maintain the quantization condition $\oint_{C}\nabla S\cdot d\mathbf{q}=nh$
on the \emph{zbw} particle, while maintaining that the diffusion of
the \emph{zbw} particle through the ether is conservative, it must
know to compensate for the phase shift experienced by the \emph{zbw}
particle passing around the solenoid along $P_{1}$, by correspondingly
shifting phase in the region that's spatio-temporally varying along
$P_{2}$. How exactly this works (assuming the ZSM framework is correct)
will presumably require developing an explicit physical model of the
ether, the \emph{zbw} particle, and the dynamical coupling of the
two, in accord with postulates (i-iii). This is left for future work.

\subsubsection{Comparison to semiclassical approximations in de Broglie-Bohm theory}

As we've seen, there does not appear to exist an empirically viable
formulation of dBBfsc-Newton/Coulomb. Nevertheless, it is possible
to formulate semiclassical approximation schemes for the `fully quantum'
formulation of dBB Newtonian gravity/electrodynamics (hereafter, dBB-Newton/Coulomb). 

The dBB-Newton/Coulomb theory corresponds to (157-158) with $V^{int}=\hat{V}_{g,e}^{int}(\mathbf{\hat{q}}_{i},\mathbf{\hat{q}}_{j})$
(for simplicity, we neglect vector potentials). In other words the
N-particle Schr{\"o}dinger equation of dBB-Newton/Coulomb is identical
to the N-particle Schr{\"o}dinger equation of ZSM-Newton/Coulomb, when
the nonlinear correction terms predicted by the latter are neglected.
The physical interpretation, however, is different. 

In ZSM-Newton/Coulomb, the \emph{zbw} particles carry rest-mass/charge
and interact with one another through the classical gravitational/electrostatic
fields they source. In dBB-Newton/Coulomb the particles are just points
at definite locations, and $\hat{V}_{g,e}^{int}(\mathbf{\hat{q}}_{i},\mathbf{\hat{q}}_{j})$
is a potential energy function on configuration space that influences
the evolution of $\psi$ in configuration space; so, to the extent
that the particles `interact' gravitationally or electrostatically,
they only do so indirectly via the influence of $\hat{V}_{g,e}^{int}(\mathbf{\hat{q}}_{i},\mathbf{\hat{q}}_{j})$
on $\psi$ through the Schr{\"o}dinger equation (157), and the influence
of $\psi$ on the evolution of the particles through the guiding equation
(158). Thus the mean-field approximation scheme discussed in section
4 applies just as well to dBB-Newton/Coulomb. 

Another dBB-based semiclassical approximation scheme has been suggested
by Prezhdo-Brooksby \cite{Prezhdo2001} and elaborated on by Struyve
\cite{Struyve2015}. Consider, for simplicity, the dBB theory with
two-particle Schr{\"o}dinger equation

\begin{equation}
i\hbar\frac{\partial\psi\left(\mathbf{q}_{1},\mathbf{q}_{2},t\right)}{\partial t}=\left[-\left(\frac{\hbar^{2}}{2m_{1}}\nabla_{1}^{2}+\frac{\hbar^{2}}{2m_{2}}\nabla_{2}^{2}\right)+\hat{V}_{g,e}^{int}(\mathbf{\hat{q}}_{1},\mathbf{\hat{q}}_{2})\right]\psi\left(\mathbf{q}_{1},\mathbf{q}_{2},t\right).
\end{equation}
The guiding equations for each particle are again given by (158),
and the 2nd-order equations of motion are
\begin{equation}
\begin{split}m_{1}\ddot{\mathbf{q}}_{1}(t)=-\nabla_{1}\left[V_{g}^{int}\left(\mathbf{q}_{1},\mathbf{q}_{2}(t)\right)+Q\left(\mathbf{q}_{1},\mathbf{q}_{2}(t)\right)\right]|_{\mathbf{q}_{1}=\mathbf{q}_{1}(t)},\end{split}
\quad m_{2}\ddot{\mathbf{q}}_{2}(t)=-\nabla_{2}\left[V_{g}^{int}\left(\mathbf{q}_{1}(t),\mathbf{q}_{2}\right)+Q\left(\mathbf{q}_{1}(t),\mathbf{q}_{2}\right)\right]|_{\mathbf{q}_{2}=\mathbf{q}_{2}(t)},
\end{equation}
where $Q\left(\mathbf{q}_{1},\mathbf{q}_{2}\right)$ is the total
quantum potential of the two-particle system. 

Now, the conditional wavefunction for particle 1, defined as $\psi_{1}(\mathbf{q}_{1},t)=\psi(\mathbf{q}_{1},\mathbf{q}_{2}(t),t)$,
satisfies the conditional Schr{\"o}dinger equation

\begin{equation}
i\hbar\frac{\partial\psi_{1}\left(\mathbf{q}_{1},t\right)}{\partial t}=\left[-\frac{\hbar^{2}}{2m_{1}}\nabla_{1}^{2}+V_{g,e}^{int}\left(\mathbf{q}_{1},\mathbf{q}_{2}(t)\right)\right]\psi_{1}\left(\mathbf{q}_{1},t\right)+K(\mathbf{q}_{1},t),
\end{equation}
where 
\begin{equation}
K(\mathbf{q}_{1},t)=-\frac{\hbar^{2}}{2m_{2}}\nabla_{2}^{2}\psi(\mathbf{q}_{1},\mathbf{q}_{2},t)|_{\mathbf{q}_{2}=\mathbf{q}_{2}(t)}+i\hbar\frac{d\mathbf{q}_{2}(t)}{dt}\cdot\nabla_{2}\psi(\mathbf{q}_{1},\mathbf{q}_{2},t)|_{\mathbf{q}_{2}=\mathbf{q}_{2}(t)}.
\end{equation}
Correspondingly, the conditional guiding equation for particle 1 is
\begin{equation}
\frac{d\mathbf{q}_{1}(t)}{dt}=\frac{\hbar}{m_{1}}\mathrm{Im}\frac{\nabla_{1}\psi_{1}}{\psi_{1}}|_{\mathbf{q}_{1}=\mathbf{q}_{1}(t)}=\frac{\nabla_{1}S_{1}}{m_{1}}|_{\mathbf{q}_{1}=\mathbf{q}_{1}(t)},
\end{equation}
where $S_{1}=S_{1}(\mathbf{q}_{1},t)$. The Newtonian equation of
motion for particle 2 is then 
\begin{equation}
m_{2}\ddot{\mathbf{q}}_{2}(t)=-\nabla_{2}\left[V_{g,e}^{int}\left(\mathbf{q}_{1}(t),\mathbf{q}_{2}\right)+Q\left(\mathbf{q}_{1}(t),\mathbf{q}_{2}\right)\right]|_{\mathbf{q}_{2}=\mathbf{q}_{2}(t)}.
\end{equation}
The semiclassical approximation is when $m_{2}\gg m_{1}$ and $\psi$
varies slowly in $\mathbf{q}_{2}$ (compared to $\mathbf{q}_{1}$).
Then $K\approx0$ and $-\nabla_{2}Q\approx0$. In other words the
time-evolution of particle 2 depends (approximately) only on the classical
interaction potential $V_{g,e}^{int}$, evaluated at the actual position
of particle 1. And the time-evolution of particle 1 depends on $\psi_{1}$
satisfying (approximately) (165) with $K\approx0$, i.e., particle
1's effective Schr{\"o}dinger equation that takes into account the back-reaction
of particle 2 through $V_{g,e}^{int}\left(\mathbf{q}_{1},\mathbf{q}_{2}(t)\right)$.
Note that, unlike models of dBBfsc-Newton/Coulomb, this semiclassical
approximation scheme defines a consistent back-reaction between the
two particles in the following sense: the conditional wavefunction
of particle 1, in the semiclassical approximation, just corresponds
to the effective wavefunction of particle 1, for which $|\psi_{1}|^{2}$
satisfies an equivariance-like property (through the conditional quantum
continuity equation implicit in (165)), even though the (semiclassically
approximated) evolution of $\psi_{1}$ still depends on the actual
position of particle 2 through $V_{g,e}^{int}$.

By contrast, the standard QM semiclassical approximation scheme for
two interacting particles \cite{Prezhdo2001,Struyve2015} is defined
by 

\begin{equation}
i\hbar\frac{\partial\psi\left(\mathbf{q}_{1},t\right)}{\partial t}=\left[-\frac{\hbar^{2}}{2m_{1}}\nabla_{1}^{2}+V_{g,e}^{int}\left(\mathbf{q}_{1},\mathbf{\overline{q}}_{2}(t)\right)\right]\psi\left(\mathbf{q}_{1},t\right),
\end{equation}
\begin{equation}
m_{2}\ddot{\overline{\mathbf{q}}}_{2}(t)=\int_{\mathbb{R}^{3}}d^{3}\mathbf{q}_{1}|\psi(\mathbf{q}_{1},t)|^{2}\left[-\nabla_{2}V_{g,e}^{int}\left(\mathbf{q}_{1},\mathbf{q}_{2}\right)\right]|_{\mathbf{q}_{2}=\mathbf{\overline{q}}_{2}(t)},
\end{equation}
where $\psi(\mathbf{q}_{1},t)$ is a single-particle wavefunction
(as opposed to a conditional or effective wavefunction), and the back-reaction
from particle 2 on particle 1 is via the average trajectory $\mathbf{\overline{q}}_{2}(t)$
inserted into $V_{g,e}^{int}$ in (169). 

Prezhdo and Brooksby \cite{Prezhdo2001} have compared the dBB-based
semiclassical approximation scheme to this standard QM scheme, for
the case of a light particle scattering off a heavy particle, where
the heavy particle is bound to a fixed surface. They found that the
dBB scheme is superior at tracking the scattering probability as a
function of time (when compared to the exact quantum dynamics description),
in addition to being computationally simpler to implement than the
standard QM scheme. 

Struyve \cite{Struyve2015} has applied the dBB-based scheme to a
dBB version of scalar electrodynamics, as well as to a dBB version
of canonical quantum gravity under the minisuperspace approximation
\footnote{Canonical quantum gravity under the minisuperspace approximation refers
to the Wheeler-DeWitt equation $\mathcal{H}\Psi\left(h,\phi\right)=0$
(and momentum constraint $\mathcal{H}_{i}\Psi\left(h,\phi\right)=0$),
under the restriction that the 3-metric $h$ and matter field $\phi$
are homogeneous and isotropic \cite{HollandBook1993,Kiefer2012,Struyve2015}.
This corresponds to a time-dependent homogeneous matter field $\phi(t)$
in an FLRW metric with homogeneous scale factor $a(t)$. The Wheeler-DeWitt
equation then takes the form $\left(H_{metric}+H_{matter}\right)\psi\left(a,\phi\right)=0$.
In the dBB version \cite{Struyve2015}, this latter form of the Wheeler-DeWitt
equation is accompanied by guidance equations for the field beables
$a(t)$ and $\phi(t)$, which turn out to be coupled to each other
via the phase $S$ of $\psi$. In this way, the metric and matter
field beables back-react on each other. It is worth mentioning that
the minisuperspace approximation is also referred to in the literature
as a `semiclassical' approximation; it should not be confused with
the dBB-based semiclassical approximation scheme, the latter of which
is applied by Struyve \emph{on top of} the minisuperspace approximation. }. In the latter case, he has compared the dBB-based scheme to the
standard scheme (applied to standard canonical quantum gravity under
the minisuperspace approximation) for cases involving macroscopic
superpositions of two Gaussians wavepackets. As it turns out, the
dBB-based scheme yields better agreement with the exact dBB version
of canonical quantum gravity under the minisuperspace approximation,
than does the standard scheme \footnote{Struyve did not compare the standard scheme to the standard quantum
interpretation of the Wheeler-DeWitt equation, the reasoning being
that the ``problem of time'' makes the standard quantum interpretation
of the Wheeler-DeWitt equation incoherent. Nevertheless, Struyve pointed
out that for approaches to quantum theory that associate approximately
classical dynamics to macroscopic superpositions of Gaussian states
(such as many-worlds interpretations \cite{Kiefer2012}), the standard
scheme is expected to do worse than the dBB scheme in approximating
exact solutions of the Wheeler-DeWitt equation (assuming those non-dBB
approaches to quantum theory yield consistent quantum interpretations
of the Wheeler-DeWitt solutions in the first place).}. 

The dBB semiclassical approximation scheme for two interacting particles
can, of course, be imported into ZSM-Newton/Coulomb. In this sense,
the results obtained by Prezhdo-Brooksby are also results that follow
from ZSM-Newton/Coulomb. However, in ZSM-Newton/Coulomb, we also have
the option of implementing the back-reaction from particle 1 onto
particle 2 via the conditional stochastic differential equation for
particle 1: $d\mathbf{q}_{1}(t)=\left(\mathrm{Im}+\mathrm{Re}\right)m_{1}^{-1}\hbar\nabla_{1}\ln\psi_{1}|_{\mathbf{q}_{1}=\mathbf{q}_{1}(t)}dt+d\mathbf{W}(t)$.
Since the trajectories predicted by this stochastic differential equation
differ from the trajectories predicted by the conditional guidance
equation (167), we would expect differences in the predictions of
the ZSM-Newton/Coulomb version as compared to the dBB version. Although,
considering that the semiclassical approximation requires the mass
of particle 2 to be much greater than particle 1, we would expect
any differences to be very slight. Nevertheless, it would be interesting
to revisit the cases studied by Prezhdo-Brooksby and Struyve, to see
if the differences might be amenable to experimental/observational
discrimination. (Revisiting Struyve's analyses from the viewpoint
of ZSM will of course require extending ZSM to relativistic field
theories in flat and curved spacetimes, and to the spacetime metric
itself. Future work will show how this can be done.)

\section{Conclusion}

We have shown how to formulate fundamentally-semiclassical Newtonian
gravity/electrodynamics based on stochastic mechanics in the ZSM formulation.
In addition, we have shown that ZSM-Newton/Coulomb has a consistent
statistical interpretation, recovers the standard exact quantum description
of matter-gravity coupling as a special case valid for all practical
purposes (even though gravity remains fundamentally classical in the
ZSM approach), and recovers the SN/SC and stochastic SN/SC equations
as mean-field approximations. We have also compared ZSM-Newton/Coulomb
to theories of semiclassical Newtonian gravity based on standard quantum
theory, dynamical collapse theories, other possible formulations of
stochastic mechanics, and the dBB pilot-wave theory. In doing so,
we have highlighted conceptual and technical advantages entailed by
ZSM-Newton/Coulomb, and indicated possibilities for experimentally
testable differences. 

In Part II, we will use ZSM-Newton/Coulomb to formulate a new `large-N'
prescription that makes it possible to consistently describe large
numbers of identical (ZSM) particles \emph{strongly interacting} classical-gravitationally/electrostatically.
This new large-N prescription will also make it possible to recover
classical Newtonian gravity/electrodynamics for macroscopic particles,
as well as classical Vlasov-Poisson mean-field theory for macroscopic
particles weakly interacting gravitationally/electrostatically.

We wish to emphasize once more the two key results of the present
paper: (i) while ZSM-Newton and ZSM-Coulomb treat the gravitational
and Coulomb potentials, respectively, as fundamentally classical fields
sourced by point-like classical particles undergoing non-classical
(stochastic mechanical) motions in 3-space, these semiclassical theories
nevertheless recover the standard quantum descriptions of Newtonian/non-relativistic
gravitational/Coulombic interactions between particles; and (ii) the
large N limit scheme of Golse \cite{Golse2003} and Bardos et al.
\cite{BardosGolseMauser2000,BardosErdosGolseMauserYau2002}, applied
to ZSM-Newton/Coulomb, makes it possible to recover the same mean-field
approximations as obtained from standard Newtonian PQG/SCEG and standard
non-relativistic QED/SCRED (the SN/SC and stochastic SN/SC equations).

In a forthcoming standalone paper, we will elaborate on one of the
possibilities for experimentally testing ZSM-Newton vs. other theories
of semiclassical Newtonian gravity, namely, the ``grav-cat'' setup
proposed by Derakhshani, Anastopoulos, and Hu \cite{DerAnaHu2016,DerakProbingGravCat2016}.
In another standalone paper, we will show how to consistently incorporate
gravitational and electrodynamical radiation reaction effects within
ZSM-Newton and ZSM-Coulomb, respectively, through a stochastic mechanical
generalization of Galley's variational principle for nonconservative
systems \cite{Galley2013}. Further down the road, we will show how
to extend ZSM to particles and fields in relativistic spacetimes,
and then use that framework to formulate consistent hidden-variables
theories of semiclassical Einstein gravity and semiclassical relativistic
electrodynamics; we will then show that the Newtonian limits of these
two theories yield ZSM-Newton and ZSM-Coulomb, respectively.

\section{Acknowledgments}

I thank Guido Bacciagaluppi, Roderich Tumulka, Bei-Lok Hu, and Dieter
Hartmann for helpful discussions.

\bibliographystyle{unsrt}
\bibliography{ZSM-Newton}

\end{document}